\documentclass[twocolumn,amsmath,amssymb,footinbib,superscriptaddress,floatfix,aps,longbibliography,prapplied]{revtex4-2}
\pdfoutput=1

\usepackage[utf8]{inputenc}
\usepackage[dvipsnames]{xcolor}
\usepackage{graphicx}
\usepackage{epsfig}
\usepackage{epstopdf}
\usepackage{amsfonts}
\usepackage{import}
\usepackage{amstext,amsmath,amssymb}
\usepackage{svrsymbols}
\usepackage{lipsum}
\usepackage{microtype}
\usepackage[none]{hyphenat}

\usepackage{rotating}
\usepackage{dcolumn}
\usepackage{array,multirow}
\usepackage[english]{babel}
\usepackage[export]{adjustbox}
\usepackage{braket}
\usepackage[caption=false]{subfig}
\usepackage{xr}
\usepackage{ifpdf}
\usepackage{tikz}
\usetikzlibrary{arrows}
\usepackage{ulem}

\usepackage{hyperref}
\hypersetup{urlcolor=violet, citecolor=violet, linkcolor=violet, colorlinks=true}
\newcommand{\be}{\begin{equation}}
\newcommand{\ee}{\end{equation}}
\newcommand{\bea}{\begin{eqnarray}}
\newcommand{\eea}{\end{eqnarray}}

\newcommand{\at}[2][]{#1|_{#2}}
\newcommand{\barr}{\vec{r}}
\newcommand{\barro}{\vec{r}_0}
\newcommand{\tev}{\tau_{\text{ev}}}
\usepackage{color}

\def\brms{B_\text{rms}}
\def\frms{\Phi_\text{rms}}
\def\tev{\tau_\text{ev}}

\DeclareMathOperator{\sinc}{sinc}

\makeatletter
\let\cat@comma@active\@empty
\makeatother

\begin{document}

\title{Models of liquid-sample confinement for nanoscale NMR}

\author{Santiago Oviedo-Casado}\email{santiago.oviedo@upct.es}
\affiliation{Área de Física Aplicada, Departamento de Física Aplicada y Tenología Naval, Universidad Politécnica de Cartagena member of European University of Technology EUT+, 30202 Cartagena, Spain}
\affiliation{Escuela Superior de Ingeniería y Tecnología, Universidad Internacional de La Rioja, 26006, Logroño, La Rioja, Spain}

\author{Daniel Cohen}\email{daniel.cohen7@mail.huji.ac.il}
\affiliation{Racah Institute of Physics, Hebrew University of Jerusalem, 91904 Jerusalem, Israel}

\author{Allan Josué González-Villalobos}\email{allan.gonzalezvillalobos@ucr.ac.cr}
\affiliation{Escuela Superior de Ingeniería y Tecnología, Universidad Internacional de La Rioja, 26006, Logroño, La Rioja, Spain}
\affiliation{Instituto de Investigaciones en Ingeniería, Universidad de Costa Rica, San Pedro, Montes de Oca, San José, Costa Rica}

\author{Javier Cerrillo}\email{javier.cerrillo@upct.es}
\affiliation{Área de Física Aplicada, Departamento de Física Aplicada y Tenología Naval, Universidad Politécnica de Cartagena member of European University of Technology EUT+, 30202 Cartagena, Spain}

\begin{abstract}
Diffusion is a prominent source of noise affecting nuclear magnetic resonance at the nanometer scale (nano-NMR), preventing high-resolution studies of unpolarized liquid samples and hindering the development of nanofluid devices. Actively managing diffusion noise through sample confinement is likely to unveil alternative noise sources that have so far been disregarded, as they occur in longer timescales and consequently are masked by diffusion. These secondary noise sources could diminish the advantages provided by sample confinement, but they may also yield valuable information about the behavior of the fluid sample and its interactions with the confinement device. In this article, we study in detail two previously unaccounted noise models for confined nano-NMR, namely, surface interactions and porosity, and discuss their implications for typical nano-NMR experiments.
\end{abstract}

\maketitle

\section{Introduction}

Nuclear magnetic resonance (NMR) spectroscopy is a fundamental tool to study molecular structures and samples composition. Based on detecting the magnetic field produced by the precessing spin of specific atomic nuclei in molecules, NMR is widely employed to chemically analyze samples of relevance in fields ranging from materials science to pharmacology. Despite having driven fundamental scientific breakthroughs in the last decades, it is an accepted truth that NMR is limited by a low sensitivity that demands substantial amounts of sample to achieve the desired precision for sample characterization \cite{Lowsensitivity}. Various solutions are currently being explored to improve the sensitivity of NMR while at the same time reducing the sample size; for example, microcoils \cite{Olson1995,Olson2004}. Among them, quantum sensors offer a most promising route \cite{Cappellaro2017}. 

Quantum sensors have demonstrated a remarkable ability to outperform their classical counterparts \cite{Giovannetti2011}. Using definite quantum properties permits estimating physical parameters with a precision far beyond what is achievable in the classical limit \cite{Maccone2006}. One of the best developed quantum sensing technologies is the NMR analog, which consists of a controllable spin that can be brought in close contact with the sample, thereby allowing the sample size to be reduced at a minimal expense in sensitivity. The most prominent platform for quantum sensing NMR is the nitrogen-vacancy (N-V) center in diamond \cite{Taylor2008}, which has been shown to be capable of detecting the magnetic field produced by the nuclei in nanosized samples of molecules at ambient conditions \cite{Ermakova2013,Mamin2013,Muller2014,Ajoi2015,Lovchinsky2016,Glenn2018,Wang2019,Barton2020}. At these scales, where the detector lies at a distance of only a few nm from the sample, the statistical polarization dominates over the thermal averaging \cite{Degen2007,Reinhard2012,Herzog2014}, allowing to perform room-temperature nanoscale NMR (nano-NMR) in pristine samples.

Despite the practical advantages that they offer, quantum sensors feature their own set of limitations, the main ones being the noise affecting the quantum properties of the sensor, the noise affecting the coupling between the sensor and the sample, and the backaction that the sensor induces on the sample. The first kind can be substantially mitigated using techniques such as dynamical decoupling \cite{Viola1999,Cywinski2007,KDDref}, while backaction can be utilized, in some scenarios, to increase precision \cite{TuviaUltimate}. Coupling noise, however, which in NMR experiments is typically dominated by diffusion of the molecules in the liquid state sample, is harder to manage without actively manipulating the sample, forfeiting the intended result for which nano-NMR was designed.

\begin{figure*}
\includegraphics[width=\textwidth]{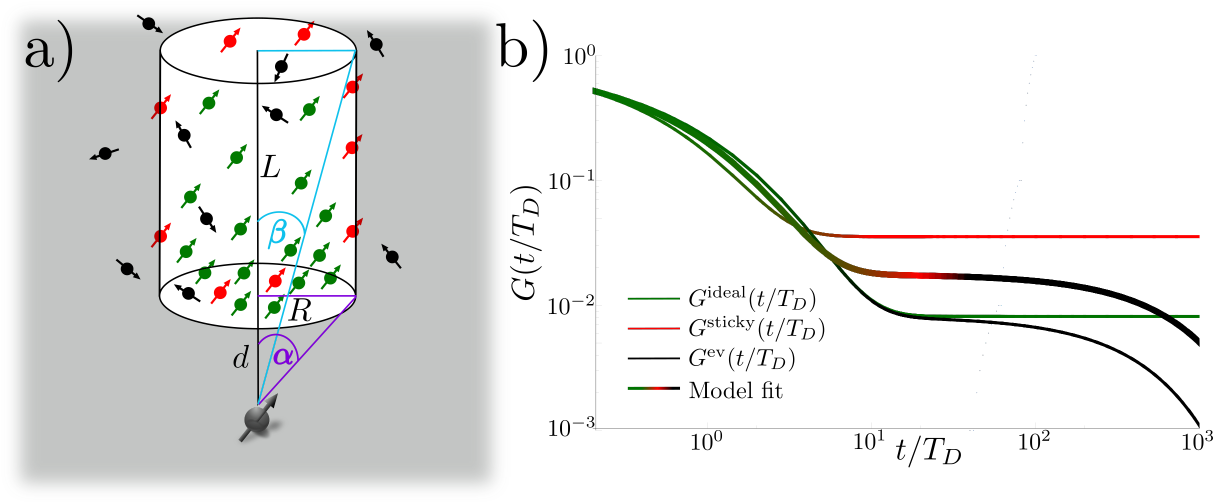}
\caption{a) Schematic depiction of the considered setup, with a cylinder volume of radius $R$ and height $L$ carved on the surface of a diamond. Inside this volume the nuclei in the sample diffuse freely. The diamond matrix features an atomic defect located at a depth $d$ below the cylinder lower cap, aligned with the symmetry axis of the cylinder. The defect position spans an angle $\alpha$ with respect to the lower border of the cylinder and an angle $\beta$ with respect to the upper border. The nuclei in the sample get statistically polarized (green) and diffuse through the confinement volume. Upon touching the walls, we consider three different conceptual model scenarios denoted by the different colors: the nuclei can rebound unchanged (green), remain attached to the surface without changing their orientation (red), alter their orientation such that they no longer contribute to the polarization signal (black), or evaporate from the confinement region (black as well). Each of these scenarios induces a distinct magnetic field at the N-V position. (b) Logarithmic scale time dependence of the correlation function for the magnetic field at the N-V position for each of the noise models considered, which represents the (postprocessed) signal that is analyzed following an experiment. Nuclei that rebound unchanged keep contributing to the overall correlations, which after some initial decay due to diffusion, then reaches a \textit{plateau} (green) in which diffusion is, effectively, stalled. Nuclei that get stuck to the walls keep contributing to the overall signal, yielding a \textit{plateau} amplitude (red) that can be significantly larger than when uniformly distributed on the inside of the cylinder for certain cylinder parameters. Nuclei that lose their orientation or seep out of the confinement volume produce a decrease of the correlation signal (black). The overall behavior of the signal correlation function is likely to be a combination of these models, as shown by the thick, multicolored line, which corresponds to a simple model for the correlations, described in Eq.\,\eqref{longtimedissipation}, that includes all of the behaviors discussed in this article.}\label{Fig1}
\end{figure*}

The problem of coupling noise stems from the dipole-dipole interaction, which is the main coupling between the nuclei in the sample and the spin sensor in nano-NMR. The dipole-dipole coupling is strongly distance-dependent, resulting in a detected signal to which only the closest nuclei contribute significantly. Diffusion yields a characteristic timescale for the statistically polarized nuclei to become, on average, too distant from the sensor to generate a signal of amplitude larger than the background noise, at which point no meaningful information can be gained from the sample. Thus, diffusion limits the precision with which the sample can be studied. On the other hand, the random nature of diffusion transfers to the couplings between nuclei and sensor, which fluctuate accordingly. Using noise spectroscopy techniques allows us to elucidate the nature of these fluctuations and, in that way, study the underlying diffusion model \cite{Mitra1992,Alvarez2011}. We propose that this knowledge can in principle be utilized to identify physical processes occurring in the sample, manage the impact of diffusion in spectroscopy, and potentially lead to increased precision.

A possible route to eliminate the problem of diffusion noise is to confine the sample to an enclosed volume of size comparable to that of the sensing region \cite{Cohen2020,Liu2022}. In the confined model, nuclei, rather than simply diffusing away from the sensor, remain, on average, within a distance defined by the size of the confining volume, resulting in an ideal steady state in which the signal is no longer affected by diffusion, and from which significant properties of the sample can be estimated with increased precision. This model, however, disregards other noise sources whose timescale is longer than that of diffusion, which become dominant once diffusion is mitigated. Among these, we find nuclei relaxation \cite{Hubbard1963} and surface interactions \cite{Wrachtrup2015,liu2023surface}, all of which mean that, ultimately, spectroscopic sensitivity is limited, but also that valuable information about the behavior of the sample components can be extracted by comparing with the appropriate noise models. 

In this article, we tackle the problem of understanding realistic confinement in a nano-NMR setting, with noise sources other than diffusion affecting the ability to extract information from the sample. To do so, we consider the analytic solution to the diffusion equation in a confined region of the space, in which we conceptualize two different mechanisms of signal decay other than diffusion, which, for clarity, we treat separately. In the first case, we model the nuclei transitioning from a confining volume to a confining (trapping) surface at a rate dictated by diffusion. We name this mechanism \textit{sticky walls}, as it describes the nuclei being immobilized upon arrival at the confining volume walls, as illustrated in Fig.\,\ref{Fig1}a), effectively reducing the diffusion coefficient to zero due to, for example, a mild electromagnetic interaction. For the second scenario, we consider a confining volume from which nuclei evaporate (dissipate) at a given, fixed rate. With this model, which we call \textit{evaporating walls}, we aim to describe, through a heuristic model, orientation changes in the nuclei due to interaction with the walls or porous confining volumes, as shown in Fig.\,\ref{Fig1}a), that lead to a randomization of the interactions and, thereby, the decay of the statistical polarization.

With the proposed models, we shed a first light on various noise processes occurring in nanosized confined fluids, and the way they get imprinted as different regimes in the detected signal, which in this setting corresponds to the autocorrelation of the magnetic field created by the statistically polarized nuclei at the probe's position, as displayed in Fig.\,\ref{Fig1}b), which can then be used to infer significant properties about the sample and its interactions with the quantum sensor environment. The information provided by the specific shape of the autocorrelation, together with the theoretical analysis that we present, can then be taken advantage of to decide the optimal experimental parameters and to improve the experiment design. As well, we demonstrate how the confinement regime can lead to enhanced parameter estimation in nano-NMR spectroscopy. Moreover, our analysis provides realistic scenarios with which to compare whether indeed confining a sample is a reasonable strategy in terms of spectroscopic precision, for different regimes.

The article is organized as follows: first, we describe our imagined setting for confined nano-NMR, and explain the procedure to calculate the correlation functions, which we then explicitly derive for the different noise mechanisms considered, providing the necessary mathematical details for their calculation. Afterward, we numerically study the behavior of the correlations for different confinement volumes and noise model settings, and analytically derive simple expressions to understand the physics of said behavior. Finally, we demonstrate the capabilities of a confined experiment in terms of spectral resolution.


\section{Methodology}

Our aim is to understand the properties and behavior of a confined distribution of molecules, through the interaction of certain nuclei in the molecules with an electronic spin. The setting is a shallow N-V center in a diamond, namely, an N-V center that is located at a depth $d$ of a few nm below the diamond surface and, consequently, in close contact with the sample sitting above that surface. We conceive that the confining volume has a cylindrical geometry of volume $V$, with the N-V center aligned with the $z$ symmetry axis of the cylinder, as shown in Fig.\,\ref{Fig1}a).

The N-V center can be treated as a spin-1/2 two-level system ($\vec{S}$) that interacts with a nucleus ($\vec{I}$) in the sample through the dipole-dipole Hamiltonian
\be
\frac{H_{DD}}{\gamma_e \Upsilon} = \frac{3(\vec{S}\cdot\hat{r}) (\vec{I}\cdot\hat{r})-\vec{S}\cdot\vec{I}}{|\barr|^3},
\ee
where $\gamma_{e}$ is the gyromagnetic ratio of the electron and $\Upsilon = \hbar \mu_0\gamma_N/4\pi$, with $\gamma_{N}$ the gyromagnetic ratio of the nuclei. $\vec{r}$ is the distance between the N-V center and each nuclei and the Hamiltonian is given in units of frequency. The $|r|^{\,-3}$ dependence defines the characteristics of the interaction, as distant nuclei hardly contribute to the overall detected signal, which corresponds to the time evolution of an initial superposition state of the N-V center interacting with the nuclei in the sample. Then, a measurement of the N-V center state at a given time reflects an instantaneous realization of the magnetic field that is produced by the statistically polarized nuclei, which we model
as $B(t) = \sum_i a_i(t)\cos(\omega_i t) + b_i(t)\sin(\omega_i t)$, where the amplitudes $\left\{a_i(t),b_i(t)\right\}$ fluctuate according to the noise affecting the sample, and $\omega_i$ are the frequencies at which the nuclei precess. For fast diffusing samples, noisy fluctuations change the amplitudes faster than the time it takes to perform a measurement, so to appropriately study the sample the optimal procedure is to target the autocorrelation of the magnetic field $\langle B(t_0)B(t_0+t) \rangle$ \cite{Staudenmaier2023}.

The different dipole-dipole interaction components can be resolved by appropriately tuning the external offset magnetic field applied to the N-V center probe, such that only one of them dominates the spectrum. Since all of them produce similar results in terms of N-V center spin state evolution \cite{Cohen2019}, and yield similar information about the sample, in this article we focus on the $zz$ component, or flip-flop term. Then, the autocorrelation function corresponding to an initial superposition state of the N-V center, evolving under the influence of the magnetic field created by a statistically polarized distribution of diffusing nuclei, is: 
\be
\begin{split}
 &G(t) = \langle B(t) B(0)\rangle \\& \propto \int \int f(\bar{r})f(\bar{r}_0)P(\bar{r},t|\bar{r}_0,t=0)d^3rd^3r_0,
\label{correlationintegral}
\end{split}
\ee
where $f(\barr) \propto Y^0_2(\rho,z)/\left|\barr\,\right|^{\,3}$ represents the form factor for the corresponding dipole-dipole coupling term, $Y^0_2(\rho,z)$ being the spherical harmonic, explicitly written in Eq.\,\eqref{eqSI:sphericalharmonic} on Appendix\,\ref{AppendixSticky}. $P(\bar{r},t|\bar{r}_0,t\!=\!0)$ is the propagator for the nuclei, which describes the probability of finding a given nucleus at the position $\barr$ at time $t$ provided an initial position $\barro$ at the initial time, that we denote as $t=0$. For the initial state of the sample we will consider that the nuclei are uniformly distributed over the confining volume. 
Our first task is then to obtain an expression for $P(\bar{r},t|\bar{r}_0,t\!=\!0)$, given diffusion in a finite, confining volume. 

In this article, we consider Fickian diffusion \cite{Fick1855}. Then, the equation governing the propagator  $P(\bar{r},t|\bar{r}_0,t\!=\!0)$ reads
\be
\frac{\partial P(\bar{r},t|\bar{r}_0,t=0)}{\partial t} = D \nabla^2 P(\bar{r},t|\bar{r}_0,t=0),
\label{Fickfluid}
\ee
where we assume a constant diffusion coefficient $D$. We further choose two different sets of boundary conditions to solve Eq.\,\eqref{Fickfluid}, each of them corresponding to a toy model describing a different physical configuration of confined diffusion that yields a particular, characteristic correlation function, and compare each of them to the ideal model of perfectly reflective walls \cite{Cohen2020}.

\section{Mathematical derivations}

This section describes the different confinement models that we consider and sketches the mathematical calculations that lead to the correlation functions for each of them, highlighting the relevant equations necessary to understand the results that follow. Additional mathematical details can be found in Appendixes\,\ref{AppendixSticky} (for the sticky model) and \ref{AppendixEvaporation} (for the evaporation model). In all cases, our starting point is the Fickian fluid diffusion equation for the propagator $P(\barr,t|\barro,t\!=\!0)$ in Eq.\,\eqref{Fickfluid} written in cylindrical coordinates:
\be\label{eq:cylindricaldiffusion}
\frac{\partial P}{\partial t} -D\left[ \frac{1}{\rho}\frac{\partial }{\partial \rho}\left(\rho\frac{\partial P}{\partial \rho}\right) + \frac{1}{\rho^2}\frac{\partial ^2P}{\partial \phi^2} + \frac{\partial ^2P}{\partial z^2}\right]=0.
\ee
The difference between the models translates to different boundary conditions with which the propagator needs to be solved. Note that in all cases we consider the same initial conditions: That is, in both models the propagator is defined with a point-like initial condition at $\barro$, satisfying $P(\barr,t=0|\barro,t=0) = \delta(\barr-\barro)$ with $\barr$ and $\barro$ restricted to $\rho,\rho_0 \in [0,R]$ in the radial coordinate and $z,z_0 \in [d,d+L]$ in the axial coordinate. The uniform initial distribution of molecules is then recovered by integrating the propagator over all initial positions $\barro$ in the confining volume.

\subsection{Sticky walls}

In the sticky model, we consider that, upon interaction with the walls, the molecules in the sample get attached to the surface of the confinement volume, such that the initial uniform distribution of molecules over the volume ends up distributed over its inner surface, from where the nuclei keep contributing to the overall statistical polarization, as shown in Fig.\,\ref{Fig1}. The sticky walls model reproduces an extreme case of stick-drag or space dependent diffusion models \cite{Kuldova2015,Somrita2020,Boehm2021,Alexandre2022}, and it is justified by studies suggesting that, near surfaces, fluids tend to staticity \cite{Vilquin2023}, experiencing an abrupt transition from the diffusion dynamics happening in the bulk.

We want to calculate the correlation function behavior as the molecules diffuse through the bulk of the confinement volume and become attached to its walls. To model this scenario we must consider Dirichlet boundary conditions, which read 
\be
\begin{split}
&P(\barr,t|\barro,t = 0) \at[\bigg]{\rho = R} = 0, \,\,\,\,  P(\barr,t|\barro,t = 0)\at[\bigg]{z = d} = 0, \\&
 P(\barr,t|\barro,t = 0)\at[\bigg]{z = d+L} = 0, \,\,\,\, P(\phi+2\pi) = P(\phi),
\end{split}
\ee
and correspond to a vanishing propagator at the cylinder boundaries. Considering these boundary conditions, the initial condition for the propagator, and aiming for an axially symmetric description, obtaining the full propagator is an exercise in partial differential equations, details of which can be found in, e.g., Refs.\,\cite{morse1953,carslaw1959,crank1975}. The full propagator expression can be adapted for the particular conditions of the problem from the propagators in Ref.\,\cite{Diffbook}, and for completeness we write it in Eq.\,\eqref{eqSI:stickypropagator} on Appendix\,\ref{AppendixSticky}. 

From the propagator, we can obtain the time evolution of the autocorrelation function inside the confinement volume. Following Eq.\,\eqref{correlationintegral}, we note that the integral over the angular coordinates $\phi$ and $\phi_0$, together with the spherical harmonics normalization conditions and the cylinder symmetry, yields $4\pi^2\delta_{n,k}$, with $k$ the specific dipole-dipole interaction term on resonance considered, which for us, being the flip-flop (or ZZ) term, is $k=0$, which means that only the $n = 0$ eigenmode for the angular coordinate survives. Considering the symmetry of the propagator, and that of the dipole-dipole interaction, the autocorrelation, normalized to the interaction strength $\Upsilon$, can be written in a rather compact form as
\be
\begin{split}\label{eq:bulkstickycorrelation}
&G_{\text{bulk}}^{\text{sticky}}(t) = \frac{8\pi^2\chi^2}{V}\sum_{m=1}^\infty\sum_{p=0}^\infty \\& \Bigg( \exp\left[-\frac{t}{\tau_{m,p}^{\text{even}}}\right]   \Bigg\{\int_d^{d+L}dz\int_0^R d\rho  \frac{\rho Y_2^0(\rho,z)}{\left(\rho^2 + z^2 \right)^{3/2}} \\& \times\frac{J_0(\beta_{p}\rho)}{J_{1} (\beta_{p}R)}\sin\left[\frac{2m\pi}{L}\left(z-d-\frac{L}{2}\right)\right] \Bigg\}^2 \\& +   \exp\left[-\frac{t}{\tau_{m,p}^{\text{odd}}}\right] \Bigg\{\int_d^{d+L}dz\int_0^R d\rho  \frac{\rho Y_2^0(\rho,z)}{\left(\rho^2 + z^2 \right)^{3/2}} \\& \times\frac{J_0(\beta_{p}\rho)}{J_{1} (\beta_{p}R)} \cos\left[\frac{(2m-1)\pi}{L}\left(z-d-\frac{L}{2}\right)\right]\Bigg\}^2\Bigg),
\end{split}
\ee
where the decay modes are defined as $\tau_{m,p}^{\text{even}} = 1/D\left(\beta_p^2 + \frac{4m^2\pi^2}{L^2}\right)$ and $\tau_{m,p}^{\text{odd}} = 1/D\left(\beta_p^2 + \frac{(2m-1)^2\pi^2}{L^2}\right)$, and we omit writing $n=0$ everywhere. Note that $\chi$ is the nuclei density which henceforth we take as $\chi = 1$ and also omit it.  The radial eigenvalues $\beta_{p}$ that correspond to the Bessel harmonics of the first kind, the functions $J_0(\beta\rho)$, are the positive solutions to the equation 
\be\label{eq:transsticky}
J_0(\beta R) = 0.
\ee

We are interested in the full time evolution of the correlation function. Molecules diffuse in the bulk following Eq.\,\eqref{eqSI:stickypropagator} and as they touch the walls, they adhere to that end position and continue contributing to the correlation function. Then, at times much longer than the so-called \textit{volume time} $\tau_V\sim\frac{V^{2/3}}{D}$ that marks the transition from a diffusive regime to a confined regime in which the walls influence the evolution \cite{Cohen2020}, we can assume that all molecules are distributed on the surface, and we can calculate the asymptotic value of the correlation function exactly through the harmonic measure $\omega(\Gamma,\barro)$ \cite{GarnettMarshall2005}, which quantifies the probability that a given molecule starting at some random position $\barro$ inside bulk, which for us is the cylinder, ends at a particular point in the surface $\Gamma$ of that region. Integrating over the whole surface of interest we get the final distribution of nuclei on that surface. The harmonic measure corresponds to the outward derivative of the Green's function in each of the surfaces of the cylinder \cite{Chaigneau2023}. The Green function in turn is the time integral of the propagator. Then, for each of the relevant surfaces $S$ on the cylinder, namely the top cap, the bottom cap, and the lateral wall, we have that 
\be
\omega^S(\barr|\barro) = D\frac{\partial}{\partial n} \int_0^\infty P(\barr,t|\barro,t=0)  dt,
\ee
and then, integrating out the angular coordinate for the corresponding spherical harmonic, the autocorrelation due to the final distribution of molecules on the surfaces of the cylinder is:
\be
\begin{split}\label{eq:stickyplateau}
&G(t\gg\tau_V) = \frac{8\pi^2}{V}\sum_{m=1}^\infty \sum_{p=0}^\infty \int_d^{d+L}dz_0\int_0^R d\rho_0 \frac{\rho_0Y_2^0(\rho_0,z_0)}{\left(\rho_0^2 + z_0^2 \right)^{3/2}} \\& \times\left\{
\int_d^{d+L} dz \frac{RY_2^0(R,z)}{\left(R^2 + z^2 \right)^{3/2}}\omega_{m,p}^{\text{wall}}(R, z|\rho_0,z_0)   \right.\\& \left. + \int_0^R d\rho \frac{\rho Y_2^0(\rho,d)}{\left(\rho^2 + d^2 \right)^{3/2}}  \omega_{m,p}^{\text{bottom}}(\rho,d|\rho_0,z_0) \right.\\& \left.
+ \int_0^R d\rho \frac{\rho Y_2^0(\rho,d+L)}{\left[\rho^2 + (d+L)^2 \right]^{3/2}} \omega_{m,p}^{\text{top}}(\rho, d+L|\rho_0,z_0)   \right\},
\end{split}
\ee
where each of the harmonic density modes reads:
\be
\begin{split}\label{eq:hdwalls}
&\omega_{m,p}^{\text{wall}}(R, z | \rho_0,z_0) = \frac{\beta_{p}J_0(\beta_{p}\rho_0)}{J_{1} (\beta_{p}R)}  \left\{ \frac{D}{\tau_{m,p}^{\text{even}}} \right. \\& \left.  \times\sin\left[\frac{2m\pi}{L}\left(z-d-\frac{L}{2}\right)\right]\sin\left[\frac{2m\pi}{L}\left(z_0-d-\frac{L}{2}\right)\right]            \right. \\& \left.  +\frac{D}{\tau_{m,p}^{\text{odd}}}\cos\left[\frac{(2m-1)\pi}{L}\left(z-d-\frac{L}{2}\right)\right] \right. \\& \left.\times\cos\left[\frac{(2m-1)\pi}{L}\left(z_0-d-\frac{L}{2}\right)\right]   \right\},
\end{split}
\ee
\be
\begin{split}\label{eq:hdcaps}
&\omega_{m,p}^{\text{bottom}}(\rho,d|\rho_0,z_0) =  \frac{J_0(\beta_{p}\rho)J_0(\beta_{p}\rho_0) }{J_{1} (\beta_{p}R)^2} \\& \times \left\{ \frac{D}{\tau_{m,p}^{\text{even}}}\frac{(-1)^m 2m\pi }{L}  \sin\left[\frac{2m\pi}{L}\left(z_0-d-\frac{L}{2}\right)\right]        \right. \\& \left.     + \frac{D}{\tau_{m,p}^{\text{odd}}} \frac{(-1)^m (2m-1)\pi }{L}  \cos\left[\frac{(2m-1)\pi}{L}\left(z_0-d-\frac{L}{2}\right)\right]   \right\},
\end{split}
\ee
and $\omega_{m,p}^{\text{top}}(\rho,d+L|\rho_0,z_0) = -\omega_{m,p}^{\text{bottom}}(\rho,d|\rho_0,z_0) $.

To calculate the transition period of diffusing nuclei we must account for those that are still diffusing in the bulk, inside the confinement volume, and those molecules that have already reached the walls, are stuck there, and continue contributing to the correlation from that final position. The former contribution is calculated through Eq.\,\eqref{eq:bulkstickycorrelation}, while for the latter we need the time-dependent accumulation on the surface as a function of time \cite{Scher2023}, which in this case corresponds to the outward flux (or first-passage time) of nuclei at the cylinder surface, that is, the time it takes for a given molecule starting at a random point $\barro$ in the bulk to first reach the surface. Then, at each point in the time evolution, we can integrate up to that time the outward flux, and calculate the corresponding contribution to the correlation through 
\be\begin{split}
&G_{\text{surface}}^{\text{sticky}}(t) =  \int_V \frac{Y_2^0(\rho_0,z_0)}{\left(\rho_0^2 + z_0^2 \right)^{3/2}} \left\{\int_S \frac{Y_2^0(\rho,z)}{\left(\rho^2 + z^2 \right)^{3/2}} \right. \\& \left. \times\left[-D\frac{\partial}{\partial n} \int_0^t P(\rho,z,t|\rho_0,z_0,t=0)  dt \right]dS\right\} dV.
\end{split}
\ee 

The crucial point to calculate the transient contribution from the walls to the correlation is to realize that, as the propagator is a sum of exponentially decaying modes, the time integral part can be easily calculated using that the propagator is a sum of temporal modes that behaves as $P(\rho,z,t|\rho_0,z_0,t=0) = \sum_s P_s(\rho,z,t\rightarrow\infty|\rho_0,z_0,t=0)\exp(-\lambda_s t)$, yielding
\be \begin{split}
&\int_0^t P(\rho,z,t|\rho_0,z_0,t=0)  dt  \\& = \sum_s \frac{P_s(\rho,z,t\rightarrow\infty|\rho_0,z_0,t=0)}{\lambda_s}\left(1-\exp(-\lambda_s t)\right).
\end{split}\ee
Then, the outward derivative of the infinite time propagator is just the harmonic measure that we need to calculate for the long-time behavior of the autocorrelation. Putting all together, the time-dependent contribution from the surface of the cylinder is
\be \begin{split}
&G_{\text{surface}}^{\text{sticky}}(t) = -\frac{8\pi^2}{V}\sum_{m=1}^\infty \sum_{p=0}^\infty \int_d^{d+L}dz_0\int_0^R d\rho_0 \frac{\rho_0 Y_2^0(\rho_0,z_0)}{\left(\rho_0^2 + z_0^2 \right)^{3/2}} \\&
\times \left\{ \left[1-\exp\left(-\frac{t}{\tau_{m,p}^{\text{even}}}\right) \right]\int_S \frac{Y_2^0(\rho,z)}{\left(\rho^2 + z^2 \right)^{3/2}}\omega_{m,p}^{\text{even}}(S|\rho_0,z_0) \right. \\& \left. + \left[1-\exp\left(-\frac{t}{\tau_{m,p}^{\text{odd}}}\right) \right]\int_S \frac{Y_2^0(\rho,z)}{\left(\rho^2 + z^2 \right)^{3/2}}\omega_{m,p}^{\text{odd}}(S|\rho_0,z_0) \right\},
\end{split}
\ee
with $\omega(S|\rho_0,z_0)$ the even (odd) --or those with sine (cosine)-- modes, respectively, of the harmonic measures for each of the surfaces $S$ of the cylinder as written in Eqs.\,\eqref{eq:hdwalls}\,and\,\eqref{eq:hdcaps}.

Note that no analytical solution exists for the sums of modes, so both the bulk and the surface regions need to be calculated numerically. The total correlation function in this sticky walls model is 
\be
G^{\text{sticky}}(t) = G_{\text{bulk}}^{\text{sticky}}(t) + G_{\text{surface}}^{\text{sticky}}(t).
\label{fullstickycorrelation}
\ee

\subsection{Evaporating walls}

The previous model assumes that upon reaching the walls, the nuclei keep their orientation, yielding a finite statistical polarization even at long-times. With the second model that we consider now, we aim to describe a polarization decay induced by the confining volume walls. In this model, either the nuclei rebound from the walls and are dispersed in random directions \cite{Gentile2017}, thereby changing their orientation, or the walls are porous and nuclei slowly seep out of the confining volume \cite{Liu2022}. In both cases the result is a decrease in the number of aligned nuclei and, consequently, a weakening of the statistical polarization, as schematically shown in Fig.\,\ref{Fig1}. We describe this scenario with the diffusion equation in a confined cylindrical volume with evaporation boundary conditions \cite{Roger2018}. Specifically, when solving the differential equation, we allow nuclei to leave the confining volume at a constant rate $\tev$, distinct from the diffusion time.

We will again be considering the diffusion propagator $P(\barr,t|\barro,t=0)$ in cylindrical coordinates, but in this case it corresponds to the solution to the diffusion equation in Eq.\,\eqref{eq:cylindricaldiffusion} with Robin boundary conditions, which
for the radial ($\rho$) and axial ($z$) coordinates are
\be 
\begin{split}
& \left(\frac{\partial P}{\partial \rho}+ \frac{d}{D\tev} P\right) \at[\bigg]{\rho = R}=0, \\& \left(\frac{\partial P}{\partial z} - \frac{d}{D\tev} P \right)\at[\bigg]{z = d} = 0, \\&  \left(\frac{\partial P}{\partial z} + \frac{d}{D\tev} P\right)\at[\bigg]{z = d+L} = 0,
\end{split}
\ee
and the periodic boundary condition on the angular ($\phi$) coordinate is $P(\phi+2\pi) = P(\phi)$. These boundary conditions link the concentration of fluid on both sides of the boundary, with a net flow rate in the direction of lowest concentration. Since we want to consider a confined sample, we assume that the concentration outside of the cylinder (i.e. at infinity) is zero, resulting in a net flow outward at the defined rate $\tev^{-1}$, thus taking into account that, at later times, less nuclei remain in the sample and, as a consequence, evaporation slows down. Note that again we consider that nuclei are initially uniformly distributed on the bulk.

Solving the evaporation model diffusion propagator is, as the previous sticky model, an exercise in partial differential equations \cite{morse1953,carslaw1959,crank1975}, and the full expression, provided in Eq.\,\eqref{eqSI:evaporationpropagator} on Appendix\,\ref{AppendixEvaporation} can be adapted from Ref.\,\cite{Diffbook} for the particular conditions of the problem. 

The autocorrelation of the evaporation model then reads 
\be
\begin{split}
G^{\text{ev}}(t) &= \frac{8\pi^2}{V}\sum_{m=1}^\infty\sum_{p=0}^\infty \frac{\beta_{p}^2}{\left[\eta_0^2+\beta_{p}^2\right]J_0\left(\beta_{p}R\right)^2} \\& \times 
\Bigg( \frac{\exp\left[-t/\tau_{m,p}^-\right]}{1-\sinc\left(\eta_m^- L\right)}\Bigg\{ \int_d^{d+L} dz \int_0^R d\rho \\& \frac{\rho Y_2^0(\rho,z)J_0\left(\beta_{p}\rho\right)}{\left(\rho^2 + z^2 \right)^{3/2}}   \sin\left[\eta_m^- \left(z-d-\frac{L}{2}\right) \right] \Bigg\}^2  \\&  
+ \frac{\exp\left[-t/\tau_{m,p}^+\right]}{1+\sinc\left(\eta_m^+ L\right)}\Bigg\{ \int_d^{d+L} dz \int_0^R d\rho \\& \frac{\rho Y_2^0(\rho,z)J_0\left(\beta_{p}\rho\right)}{\left(\rho^2 + z^2 \right)^{3/2}}  \cos\left[\eta_m^+ \left(z-d-\frac{L}{2}\right) \right] \Bigg\}^2\Bigg),
\label{evaporationcorrelation}
\end{split}
\ee
where the eigenvalues $\eta^\pm$ for the trigonometric functions describing diffusion in the axial coordinate correspond to the solutions to the transcendental equation 
\be\label{Eq:transendental1}
\left(\frac{\eta_0}{\eta^\pm}\right)^{\pm 1} = \pm \tan\frac{\eta^\pm L}{2},
\ee
with $\eta_0 = d/D\tev$, and as before we denote $\beta_p \equiv \beta_{n=0,p}$ for simplicity, after having integrated out the angular modes as we did for the sticky model. These $\beta_{p}$ eigenvalues are, in this case, the positive solutions to the equation  
\be\label{Eq:transendental2}
J_0'(\xi) + \frac{d R}{\xi D\tev}J_0(\xi) = 0, \,\,\,\, \xi = \beta R.
\ee
Finally, the coefficients  $\tau_{m,p}^{\pm} = 1/D(\beta_p^2+\eta_m^{{\pm}^2})$ in Eq.\,\eqref{evaporationcorrelation} describe the different decay modes of the correlation function.


\section{Results}

We now explicitly calculate the correlation function in the sticky walls and evaporating walls scenarios, for different confining cylinder volumes, and various evaporation rates. In both cases, we benchmark the results by comparing with the ideal scenario in which molecules rebound from the walls of the confining volume and nuclei retain their spin orientation after the collision, such that the correlation signal detected by the N-V center follows the diffusion process up to the volume time $\tau_V$, that marks the onset of a \textit{plateau} in which the correlation remains constant in time and diffusion has no further effect on the statistical polarization of nuclei.

In this idealized limit, the correlation function can be calculated exactly by integrating a uniform distribution of nuclei in the region. Following Ref.\,\cite{Cohen2020} this exact equation is
\be
G^{\text{ideal}}(t \gg \tau_V) =  \frac{4\pi^2}{V}\left[\int_d^{d+L}dz \int_0^R d\rho \frac{\rho Y_2^0(\rho,z)}{(\rho^2 + z^2)^{3/2}}\right]^2,
\ee
whose functional form gets simplified if we consider the angular relation between the lower and upper borders of the cylinder confining region. Taking the N-V center position as the lower vertex of two triangles of sides $R$, and $d$ or $d+L$ along the cylinder axis, and whose long side is the distance between the N-V center and the borders of the cylinder lower or upper cap, we can define $\alpha$ and $\beta$ to be the angles between said long sides and the alignment axis, as shown in Fig.\,\ref{Fig1}, and whose cosines are $\cos\alpha = R/\sqrt{d^2 + R^2}$ and $\cos\beta = (d+L)/\sqrt{(d+L)^2+R^2}$. Then, the amplitude of the \textit{plateau} is 
\be
G^{\text{ideal}}(t \gg \tau_V) = \frac{4\pi^2}{V}\left(\cos\beta-\sin\alpha\right)^2.
\label{idealplateau}
\ee

Further, in both the sticky and the evaporation model, we test the autocorrelation evolution in time against that of free diffusion, which should be apparent before the volume time $\tau_V$ at which confinement effects appear. The correlation function for free diffusion can be calculated exactly \cite{Cohen2019}, yielding
\be
\begin{split}
G^{\text{free}}\left(\tilde{t}\right)&=\frac{4}{\sqrt{\pi }} \left[\tilde{t}^{-3/2}-\frac{3 }{2\sqrt{\tilde{t}}} -\frac{3 \sqrt{\pi }}{2}\tilde{t}+3 \sqrt{\tilde{t}}+\frac{\sqrt{\pi }}{4} \right. \\& \left. - \left(\tilde{t}^{-3/2}-\frac{3 \tilde{t}^{3/2}}{2}-\frac{1}{\sqrt{\tilde{t}}}+\frac{7 \sqrt{\tilde{t}}}{4}\right) \right. \\& \left. \times\sqrt{\frac{\pi }{\tilde{t}}} \text{erfc}\left(\frac{1}{\sqrt{\tilde{t}}}\right) \exp \left(\frac{1}{\tilde{t}}\right)\right],
\label{Cohenian}
\end{split}
\ee
with scaled time parameter $\tilde{t} = t/T_D$. Eq.\,\eqref{Cohenian} features two distinct regimes that can be compared to the typically assumed exponential $\exp(-t/T_D)$ behavior of diffusion induced decay in nano-NMR \cite{Walsworth2016}. The first is the early time evolution $t \ll T_D$, which approximates $1-6t/T_D + O(t^2)$ and therefore decreases faster than the heuristic exponential model. The second, corresponding to large times $t \gg T_D$, displays a decay that approximates $\sim (t/T_D)^{-3/2}$, having a markedly slower decay than the simple exponential model.

\subsection{Sticky walls}

In the sticky walls model, the nuclei that end trapped on the walls continue contributing to the overall correlation signal from that fixed position. As diffusion proceeds, eventually all nuclei will be attached to the confining volume walls. Then the correlation signal also reaches a \textit{plateau} with an amplitude that depends on the surface-to-volume (or radius to height) ratio of the confinement region through the specific nuclei distribution profile in each of the surfaces, as governed by the corresponding harmonic functions modes in Eqs.\,\eqref{eq:hdwalls}\,and\,\eqref{eq:hdcaps}.

This \textit{plateau} section of the correlation function is crucial for parameter estimation, as any signal whose decay follows said correlation profile behaves, upon reaching the flat section, as a coherent signal. It is, therefore, ideal for spectroscopy \cite{Cohen2020,Oviedo2020}. As well, the full behavior of the correlation function is interesting in itself, as it contains information about the different physical processes affecting the confined sample evolution. To calculate the time evolution of the autocorrelation for the sticky walls model, we first obtain the radial eigenvalues which are the solutions to Eq.\,\eqref{eq:transsticky}, and then numerically evaluate Eq.\,\eqref{fullstickycorrelation} calculating up to 500 modes. The reason for needing such a high number lies in the precision required for the harmonic density. The bulk autocorrelation comprises exponentially decaying contributions which are  dominated by the slowest decaying modes, namely the lowest order eigenmodes, which are the ones that yield the largest $\left\{\tau_{m,p}^{\text{even}},\tau_{m,p}^{\text{odd}}\right\}$ in Eq.\,\eqref{eq:bulkstickycorrelation}. However, the harmonic measures in Eqs.\,\eqref{eq:hdwalls})\,and\,\eqref{eq:hdcaps} feature summations of terms in which the importance of the modes decreases linearly  rather than exponentially with the mode number, meaning that for the large confinement volumes ($R,L \gg d$) that can be realized experimentally a large number of terms is required to reach convergence.

\begin{figure}
\includegraphics[width=\columnwidth]{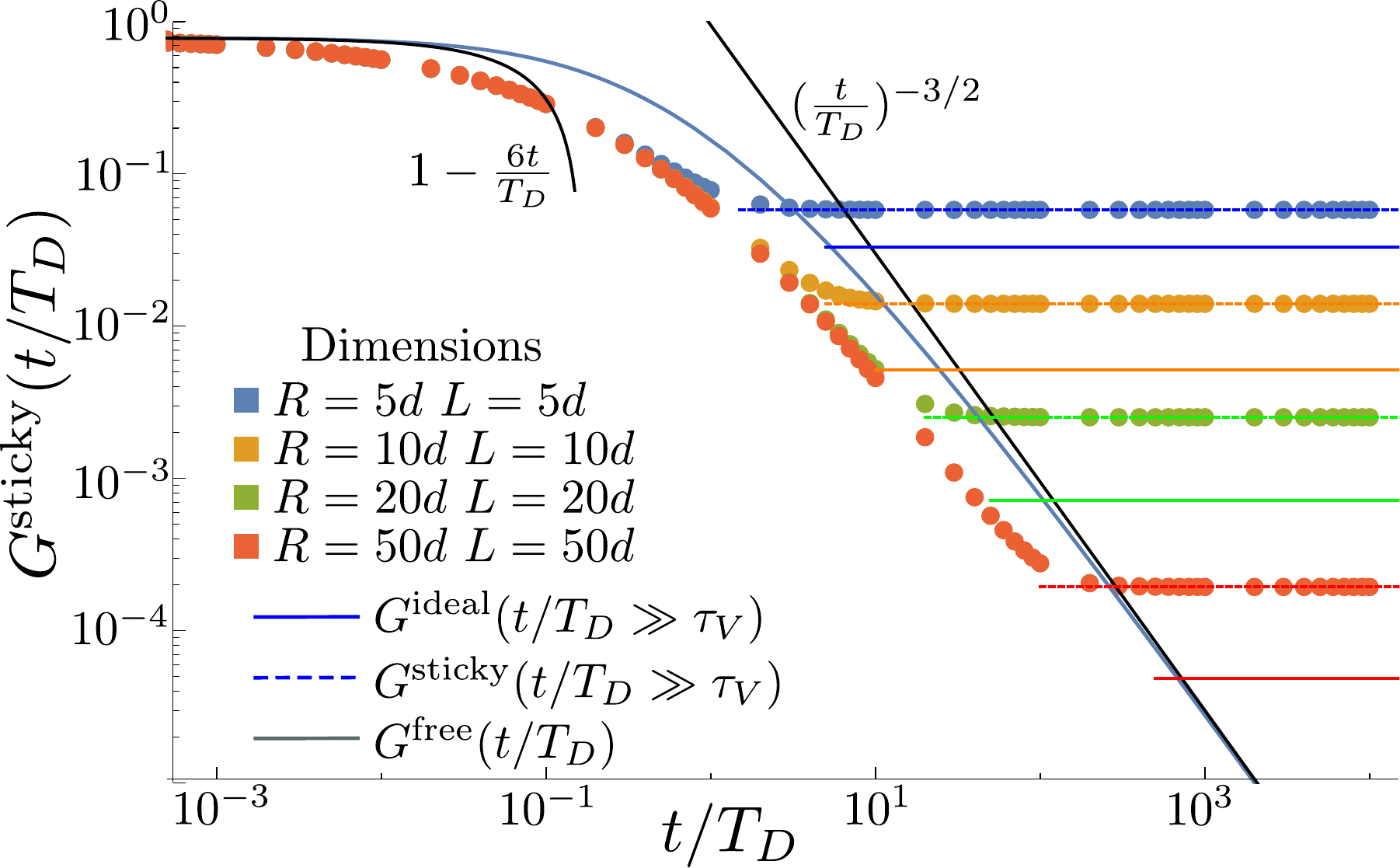}
    \caption{Logarithmic scale time evolution of the correlation function Eq.\,\eqref{fullstickycorrelation} for confined nano-NMR with sticky walls (dots), for different confinement volumes. Asymptotic behavior of the correlation function in the confinement regime for both the ideal, completely reflective walls [Eq.\,\eqref{idealplateau}, full horizontal lines, calculated exactly] and fully sticky walls [Eq.\,\eqref{eq:stickyplateau}, dashed horizontal lines]. Early time diffusive behavior and asymptotic time-dependence of the diffusive regime of the sample are displayed as solid, black lines labeled with the corresponding time-dependence, and the free diffusion behavior, described by Eq.\,\eqref{Cohenian} shown as a solid gray line. Note that we use units normalized to the N-V center depth $d$ and diffusion time $T_D = d^2/D$, to display the universal behavior.} \label{Stickycorrelation}
\end{figure}

Figure\,\ref{Stickycorrelation} displays the results for the autocorrelation function for the sticky model considering four different confinement region volumes with geometrical symmetry, meaning equal height and radius. We can see from Fig.\,\ref{Stickycorrelation} that Eq.\,\eqref{fullstickycorrelation} approximately reproduces the free diffusion early time behavior, where owing to surface accumulation instead of infinite diffusion, the free diffusion amplitude limit is not saturated. Rather, we observe a linear initial decay followed by a rapid onset of the confinement regime, which is for this geometry and model faster than what is expected in the ideal scenario of perfectly reflective walls. The power-law behavior from long-time free diffusion is only approximated for the largest confinement volumes, where the influence of the confinement boundaries is slower to appear. Of note, we can observe non-trivial deviations not only from free diffusion but also from  the ideal reflective walls scenario, with a significantly larger \textit{plateau} amplitude. Combined with the analytical formulas that we have derived, these results can be used to assess the degree of stickiness of the confinement region boundary for known geometries, or to explore geometries with controlled sticky fluids.

\begin{figure}
\includegraphics[width=8cm]{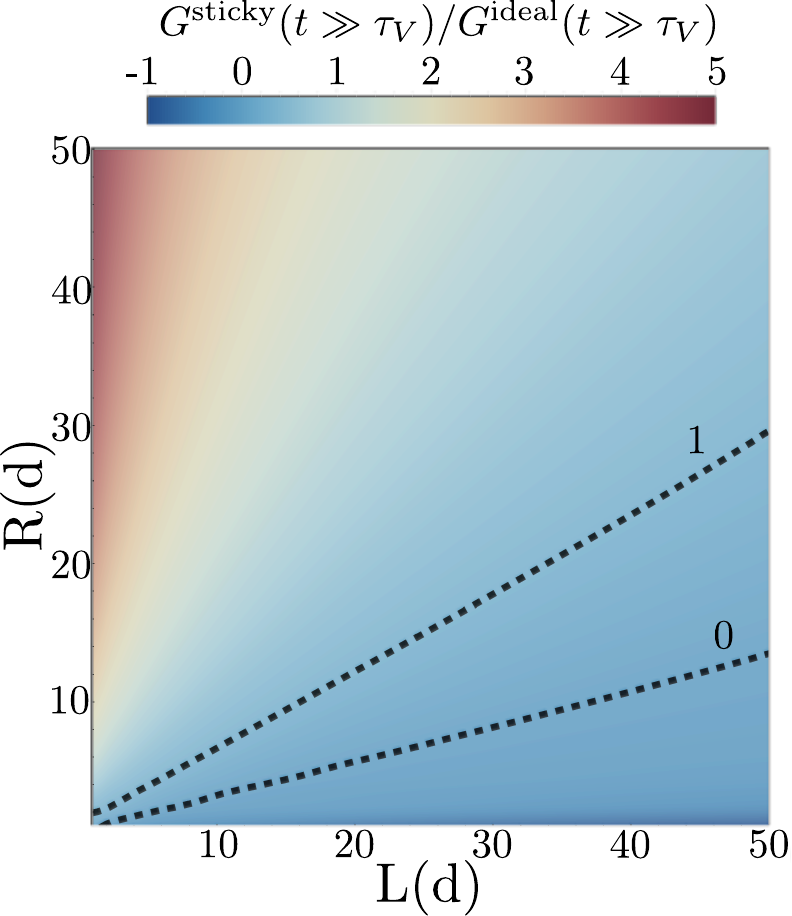}
    \caption{Signed logarithm \{sign(x)$\log_{10}[1 + \textrm{Abs(x)]}$\} \textit{plateau} amplitude ratio between correlations in the sticky and completely reflective cases, as defined by Eq.\,\eqref{plateausratio}. Here, we choose a depth $d$ = 15 nm, typical of nano-NMR experiments with N-V centers \cite{Staudenmaier2022}, and display the ratio as a function of the confinement cylinder size, defined by its radius $R$ and height $L$. We observe a marked distinction for asymmetric cylinders, with flat $R\gg L$ cylinders favoring the sticky walls model which shows a markedly larger \textit{plateau} amplitude. Elongated cylinders $L \gg R$ on the contrary favor reflective walls when graded by the \textit{plateau} amplitude, with the transition of equal amplitude (displayed by the $1$  line) located approximately at $L = 3R/2$. Note that for elongated cylinders the autocorrelation of the sticky model becomes negative (hence the signed logarithm), which in the figure is displayed by the $0$ line located around $R = L/4$, which means that the nuclei become anticorrelated with respect to the uniform initial distribution.} \label{Ratio}
\end{figure}

The long-time behavior shows the characteristic \textit{plateau} from confinement with no loss of polarization, whose amplitude is, in Fig.\,\ref{Stickycorrelation}, visibly larger than in the ideal reflective walls scenario. Both the onset time of this \textit{plateau}, defined by $\tau_V$, and its amplitude, are crucial for confined nano-NMR applications. To quantify the difference between ideal and sticky confinement, we can define the ratio between the long-time behavior of both through \be\label{plateausratio}
\frac{G^{\text{sticky}}(t \gg \tau_V)}{G^{\text{ideal}}(t \gg \tau_V)}, \ee
which we can use to estimate whether having sticky walls could prove to be, for a given confinement geometry, advantageous. Fig.\,\ref{Ratio} displays this ratio between amplitudes for significant confinement volumes.

Liquid-state nano-NMR spectroscopy greatly benefits from having autocorrelation signals whose diffusion decay is forestalled by confinement, leading to the \textit{plateau} in which the coherent signal limit is recovered. However, even if we assume that such a no-decay section of the autocorrelation is eventually reached, its amplitude is crucial, as it means the difference between a distinguishable signal that raises above the noise floor and one that gets masked by other noise sources rather than diffusion. Consequently, controlling the amplitude of the \textit{plateau} region, and accelerating its onset as much as possible is critical for successful experiments. For perfectly reflective walls, Eq.\,\eqref{idealplateau} can be used to determine the optimal confinement cylinder geometry that maximizes the amplitude, and which corresponds to $R \approx 0.93d$ and $L \approx 0.88d$. That is, a mostly symmetrical cylinder, for which we see in Fig.\,\ref{Ratio} that having sticky walls results in a larger \textit{plateau} amplitude. Note that partially sticky walls, namely a more realistic situation where nuclei both reflect on the boundary and get trapped there, would still be advantageous over the ideal optimal completely reflective walls. 

Considering the \textit{plateau} amplitude for sticky walls, and its comparison with reflective walls in Fig.\,\ref{Ratio} we observe three distinct regimes. For $R \geq L,$ the sticky walls scenario has larger amplitude, the transition occurring approximately around $R \approx 2L/3$. For flat cylinders in which $R \gg L$, meaning those approaching a two-dimensional flat confinement, the amplitude of the \textit{plateau} dramatically favors sticky walls, with an appreciable increase in the overall amplitude, being therefore a very favorable scenario for spectroscopy. On the opposite end, for deep cylinders with $L > R$, having reflective walls is a better strategy. Importantly, at $R \approx L/4 $ a transition occurs where the correlation function for sticky walls changes sign, thus reflecting that for significantly deep cylinders the final distribution of nuclei on the walls is anti-correlated with the initial uniform distribution on the bulk. While the \textit{plateau} amplitude is quite small in this region, and in fact for any $L \gg R$ for sticky walls, precluding somewhat its use in spectroscopy, the negative correlations could be of significance for studying surface or border effects in restricted diffusion settings, where having analytical solutions as those that we provide bolsters the ability to engineer specifically tailored confinement volumes.

We can rationalize the \textit{plateau} amplitude for the sticky walls model by looking at some limits in Eqs.\,\eqref{eq:hdwalls} and \eqref{eq:hdcaps} and the influence they exert on Eq.\,\eqref{eq:stickyplateau}, and comparing with the amplitude for the ideal reflective walls. First, we look at how particles distribute on each of the surfaces, namely the cylinder caps and its lateral wall, and which is controlled solely by the harmonic densities. For the caps, the Bessel harmonics determine a radial profile for particles density, meaning that closer to the z-axis, namely for $\rho \rightarrow 0$, the concentration of particles increases. For the lateral wall, it is the oscillating modes that determine a higher concentration toward the middle ($L/2$) of the cylinder. Then, the specific final amplitude is determined by these profiles and the percentage of particles that end up in each of the surfaces, which is controlled by the geometry of the cylinder. Thus, approximately, the walls get $L/(L+2R)$ while each of the caps gets $R/(L+2R)$ of the total number of particles. Considering these approximations, the $1/r^3$ dependence of the dipole-dipole interaction, and the specific flip-flop term we are focusing on, allows a qualitative explanation for the numerical results displayed in Figs.\,\ref{Stickycorrelation} and \ref{Ratio}. 

For a symmetric $L = R$ cylinder, approximately 1/3 of the particles are in each surface after a large time. Additionally, on the caps, the Bessel harmonics dictate a radial profile in which more particles concentrate toward the center, as $\rho \rightarrow 0$, this means that, in the bottom cap, a large number of statistically polarized nuclei are much closer to the sensor than what happens for uniformly distributed particles on the bulk that is the end result of perfectly reflective walls, explaining the larger amplitude for sticky walls, which is particularly striking when the two-dimensional axial confinement limit $R \gg L$ is considered, when most nuclei concentrate close to the sensor located at $\rho = 0$, rather than being uniformly distributed in the available space. On the opposite end, for elongated $L > R$ cylinders, caps lose significance getting a small number of particles, and most particles concentrate at a distance greater than $L/2$ from the sensor, which is larger than the average distance for uniform distribution on the bulk, explaining why in this geometry ideal reflective walls offer a larger amplitude. As $L$ grows to fulfill the limit $L \gg R$, lateral diffusion dominates, and most particles concentrate on the walls with a gradient concentration that is higher toward the middle $L/2$. If then we consider the specific spherical harmonic [Eq.\,\eqref{eqSI:sphericalharmonic}], we have that its sign is dictated by $2z^2 - \rho^2$, while its amplitude is governed by $\left(\rho^2 + z^2 \right)^{5/2}$. The wall of the cylinder, in which $\rho = R$, contains a small band of negative autocorrelation toward the bottom, that is for $z < R/\sqrt{2}$. Crucially, this band remains constant as the cylinder elongates with growing $L$, and for small $R$ it remains close to the probe at a distance $\propto R$ while most of the particles will be at a distance $\propto L$. Then, for dipolar interaction in which the interaction strength decreases as the inverse cube of the distance, the negative contribution dominates for $L \gg R$, particularly when the caps contribution, which scales as $1/L$, becomes negligible compared to the amount of particles the wall gets, which approaches 1. As the harmonic densities are always positive, it is the geometry of the confinement volume that determines whether the signal is correlated or anticorrelated for the sticky walls model, opening the possibility of engineering confinements tuned to specific effects.

A prime forecast application of nano-NMR is the ability to capture the dynamics of diffusion processes at the nanoscale \cite{Cohen2019}, from which the diffusion coefficient of fluids can be accurately estimated \cite{Staudenmaier2022}. Typically, the diffusion coefficient is determined by fitting the decay shape of the experimental correlation to an exponential function approximately $\exp(-t/T_D)$ \cite{Walsworth2016}, with $T_D$ defined as the lifetime of the exponential. Taking $T_D = d^2/D$, yields the diffusion coefficient $D$. Yet the exponential model is just a rough approximation for diffusion at the nanoscale, valid, as we readily observe from Fig.\,\ref{Stickycorrelation}, only at early times, where the exact correlation function envelope due to diffusion in Eq.\,\eqref{Cohenian} approximates $G^{\text{free}} (t \rightarrow 0) \approx 1-6t/T_D$, which can be taken as $\approx \exp(-6t/T_D)$ to explain the origin of the exponential model. The long-time behavior of free diffusive fluids follows a power-law, which from Eq.\,\eqref{Cohenian} is $G^{\text{free}} (t \rightarrow \infty) = 32\left(t/T_D\right)^{-3/2}/15\sqrt{\pi}$. Then, since in general the fitting is done to an imprecise model, inaccuracies in estimating $D$ occur.

On the other hand, sample confinement results in correlations which, for the sticky model (or the ideal model of perfectly reflective walls), after some characteristic time $\tau_V$, no longer decay, meaning that, on average, no diffusion is happening in the sample. The transition from an early time diffusive regime to a long-time confined regime is smooth, as shown in Fig.\,\ref{Stickycorrelation}, but given the analytical expressions derived for the harmonic measures in Eqs.\,\eqref{eq:hdwalls} and \eqref{eq:hdcaps}, the volume time $\tau_V$ that marks the transition between diffusive and confinement at the crossing time between the two regimes can be determined through
\be
\tau_V = \left[\frac{32}{15\sqrt{\pi}G^{\text{sticky}}(t\gg\tau_V)} \right]^{2/3}T_D.
\ee 
This time can be accurately estimated in experiments from the crossing point of the decaying and flat sections of the correlations. Knowing the depth of the probe, which can be independently estimated from power spectrum measurements \cite{Walsworth2016,Staudenmaier2022}, and using the definition of the diffusion time as $T_D = d^2/D$ we get that
\be\label{eq:Destimation}
D =  \left[\frac{32}{15\sqrt{\pi}G^{\text{sticky}}(t\gg\tau_V)} \right]^{2/3} \frac{d^2}{\tau_V}.
\ee

Eq.\,\eqref{eq:Destimation} offers a method to potentially improve the estimation of the diffusion coefficient $D$ of the fluid, where the main limitation lies in the accuracy of the confinement model considered. Uncertainty in the exact geometry of the confinement volume introduces a source of error which, as the technology to fabricate nanostructures in the diamond surface is well developed \cite{Birgit2010,Mehedi2013,Aiden2020}, becomes systematic, and can be partially controlled. Moreover, the successful experiment targeting the diffusion coefficient requires, contrary to the case for spectroscopy, a large confinement volume that allows for a larger diffusive regime in which the power-law behavior is more exactly achieved, and which should be easier to fabricate. Additional sources of error include surface roughness or incomplete confinement due to imperfect covering of the cylinder, which can be addressed through more accurate modeling. Note that the equations above for $\tau_V$ and $D$ are equally valid if we consider ideal reflective walls, by simply replacing $G^{\text{sticky}}$ for $G^{\text{ideal}}$ in Eq.\,\eqref{idealplateau}.

\subsection{Evaporating walls}

We now consider the scenario in which the walls cause a decay of the correlations  due to, for example, seepage of the nuclei out of the confinement volume \cite{Liu2022}, or as a result of orientation changes in the nuclei upon interaction with the walls. The correlation function in this scenario is calculated by numerically solving  Eq.\,\eqref{evaporationcorrelation}, which comprises solving two transcendental equations and calculating (truncated) infinite sums. To do so, we first numerically solve Eqs.\,\eqref{Eq:transendental1}\,and\,\eqref{Eq:transendental2} and truncate the list of solutions by calculating the decay coefficients to which each of them leads. Contrary to the sticky model, in this case the full correlation function decay is well captured by just the 100 slowest eigenmodes corresponding to the largest $\tau_{m,p}^\pm$, as higher-order modes rapidly decay in time and, therefore, have a negligible contribution to Eq.\,\eqref{evaporationcorrelation} given the considered regime. We then calculate all the relevant terms in the sum and, with them, the correlation function for different confinement volumes and dissipation rates. 

\begin{figure}
\includegraphics[width=\columnwidth]{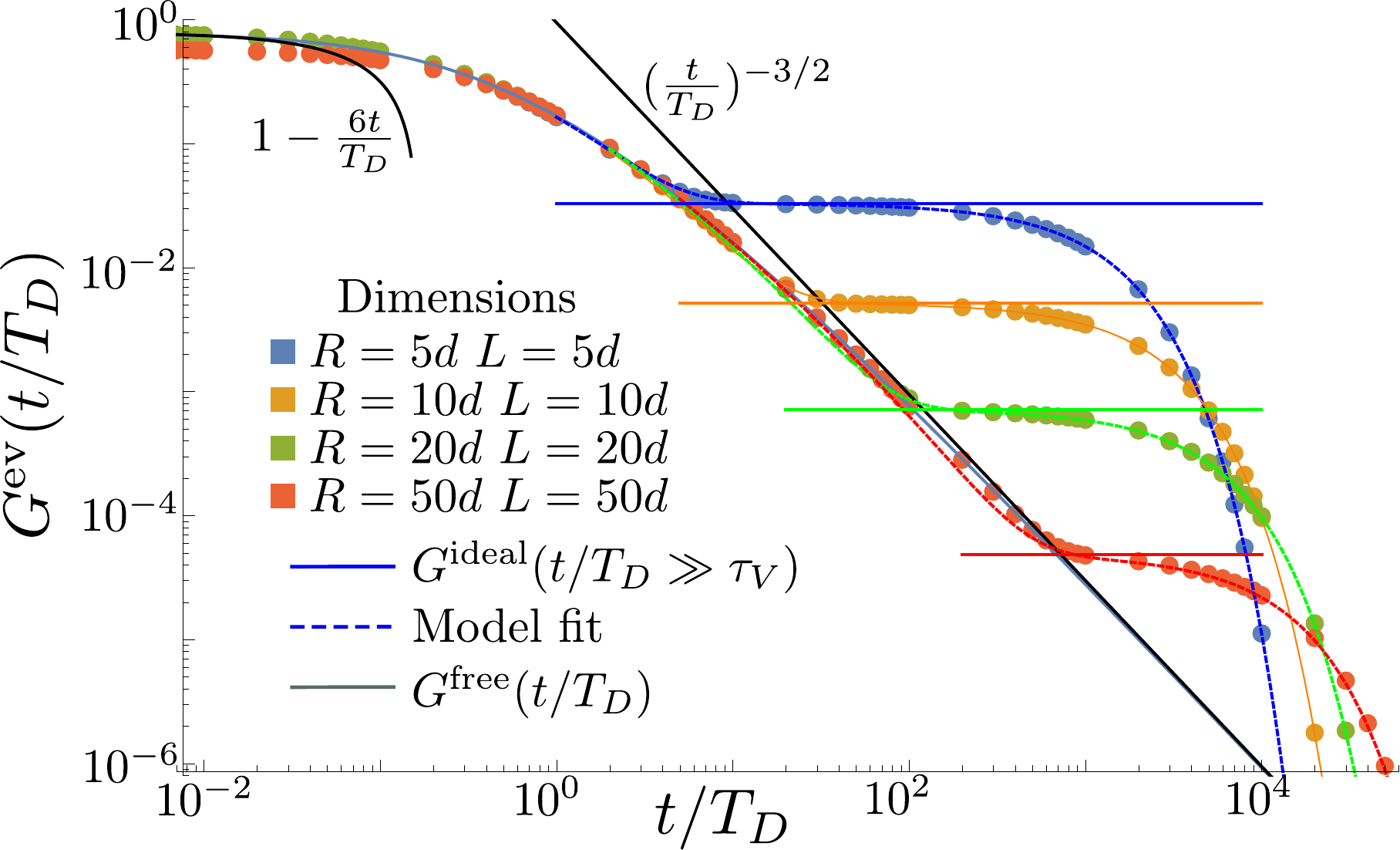}
    \caption{Logarithmic scale time evolution of the correlation function for a confinement volume with evaporating walls, calculated from Eq.\,\eqref{evaporationcorrelation}, considering $\tev = 1000T_D$ and various confinement volumes (dots). For reference, the free diffusion correlation function Eq.\,\eqref{Cohenian} is shown as a gray line on the background, and its asymptotic behavior at very short and very long times as black lines labeled with the corresponding time dependence. As well, the ideal reflective walls \textit{plateau} amplitude from Eq.\,\eqref{idealplateau} is shown as a horizontal solid line for each of the volumes considered. The model fitted to each of the correlation functions is shown as dashed lines, and corresponds to Eq.\,\eqref{longtimedissipation}, with the appropriate rates adjusted according to the different confinement volumes considered. Note that we use units normalized to the N-V center depth $d$ and diffusion time $T_D = d^2/D$, to display the universal behavior.} \label{DissipationVolumes}
\end{figure}

In Fig.\,\ref{DissipationVolumes} we display the correlation function corresponding to a fixed dissipation rate $\tev = 1000T_D$ and four different confinement volumes, together with the correlation function envelope Eq.\,\eqref{Cohenian} that describes the free diffusion scenario, and its asymptotic behavior at very short and long times. For all confinement volumes considered, we observe three distinct regions. The first one is described by free diffusion, and behaves linearly at the very beginning, showing the characteristic power-law behavior of free diffusion at later times, which is more pronounced whenever the confinement volume is bigger, as the influence of the walls is slower to appear. 

As the nuclei reach the walls of the confinement volume, they either rebound and continue contributing to the correlation, or dissipate at a given rate which, in Fig.\,\ref{DissipationVolumes}, is $\tev = 1000T_D$, and that is responsible for further decay of the correlations. When the timescales of diffusion and evaporation are markedly different, with evaporation slower than diffusion, two distinct decay regimes appear on the correlation function which are well captured by a simple model. The first comprises early time free diffusion, when the molecules have not yet explored the confinement geometry. This is followed by a long-time characteristic confinement \textit{plateau} that decays exponentially at a rate dominated by the largest coefficient $\tau_{0,0}^+$ in Eq.\,\eqref{eqSI:evaporationpropagator}, which obeys, in the slow evaporation $\tev \gg T_D$ regime, the surface-to-volume scaling $\tau_{0,0}^+ \approx Sd/V\tev$, consistent with the results for diffusion controlled surface relaxation \cite{Korringa1962,Brownstein1979,Grebenkov2007}.  To describe the transition between diffusive and dissipative-confinement regimes, the free diffusion correlation from Eq.\,\eqref{Cohenian} includes as well an exponential term that accounts for the decay of diffusion as more nuclei rebound on the walls and re-enter the sensing region rather than evaporating, thereby leading to the \textit{plateau}. Diffusion-dominated decay can be understood through a half-life for the nuclei to reach the walls, which relates to the volume time that marks the crossover between diffusion and confinement regimes as $2/\tau_V$. Then
\be
\begin{split}
G^{\text{ev}}(t) &\approx \brms^2 G^{\text{free}}(t/T_D)\exp(-\frac{2t}{\tau_V}) \\& + G^{\text{ideal}}(t \gg \tau_V )\exp\left(-\frac{Sd}{V\tev}t\right).
\label{longtimedissipation}
\end{split}
\ee
Note that the $\brms^2$ is the initial amplitude of the signal [see Eq.\,\eqref{eq:brms} in Appendix\,\ref{AppendixSticky}]. Thus, a simple model composed of two parameters accurately describes the complex dynamics of the evaporating confined fluid. Given the relation between the different decay rates and the geometry of the confinement volume, the validity of Eq.\,\eqref{longtimedissipation} increases as $\tev$ becomes larger than $T_D$, but also for more symmetrical ($R\sim L$) the cylinders which have smaller surface-to-volume ratio.

Evaporation causes a rapid degradation of correlations. Eq.\,\eqref{longtimedissipation} tells us that, whenever the dissipation time is sufficiently long compared with the diffusion time, the \textit{plateau} region characteristic of confinement will be reached before dissipation has a significant effect on the signal. On the contrary, for larger volumes, said \textit{plateau} is hardly achieved but, on the other hand, dissipation happens at a slower rate, as nuclei take longer to reach the walls. The factor $Sd/V$, that modulates $\tev$ in Eq.\,\eqref{longtimedissipation}, leads to an optimal trade-off between dissipation on the surface and diffusion on the bulk, where the design of the confinement volume can be tailored to the given dissipation rate that the walls are known to induce, to minimize its impact through maximizing the volume for a given surface, or minimizing the surface for a given volume. Moreover, it is potentially useful again to estimate the diffusion coefficient of the fluid sample, as Eq.\,\eqref{longtimedissipation} describes well the transition point between pure diffusion and the effect of the walls at $\tau_V$, much like in the case of sticky walls. Here we find again a situation where a bigger confinement yields a better estimation of $D$. It also tells us that in the case of combined effects, namely sticky and dissipative regimes combined, we can find a simple expression to describe the long-time evolution of the correlation, by replacing $G^{\text{ideal}}(t\gg\tau_V)$ with $G^{\text{sticky}}(t\gg\tau_V)$ in Eq.\,\eqref{longtimedissipation}.

\begin{figure}
\includegraphics[width=\columnwidth]{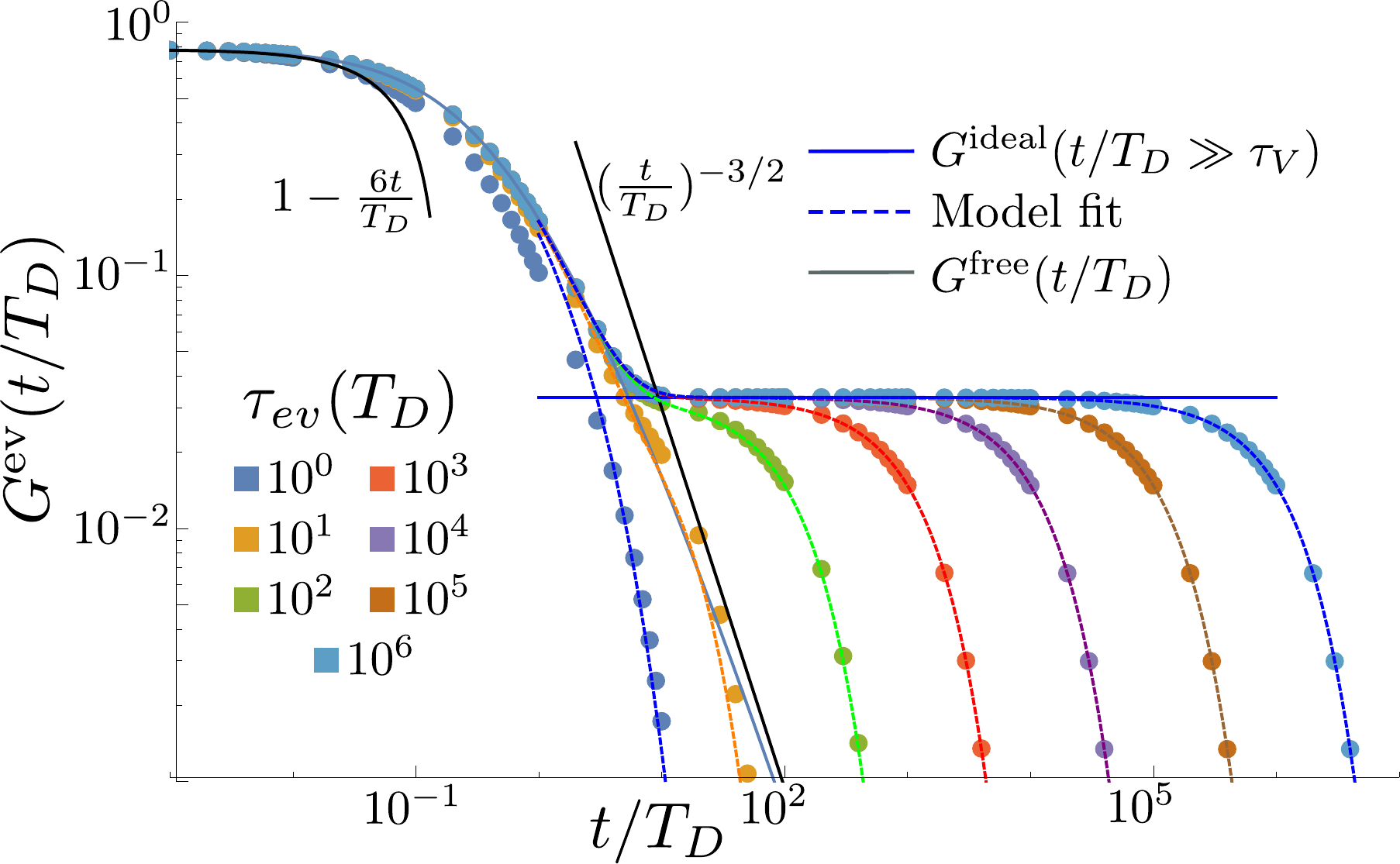}
    \caption{Logarithmic scale time evolution of the correlation function for a confinement volume with evaporating walls, calculated from Eq.\,\eqref{evaporationcorrelation} by considering a fixed confinement volume of size $R = 5d$ and $L = 5d$, and varying the evaporation rate $\tev$ in powers of ten (dots). Free diffusion correlation function Eq.\,\eqref{Cohenian} (solid gray line) and its asymptotic behavior at very short and very long times (solid black lines). Ideal reflective walls \textit{plateau} from Eq.\,\eqref{idealplateau} for the volume considered, showing that the \textit{plateau} is achieved for slow evaporation. The model fitted to each of the correlation functions corresponds with Eq.\,\eqref{longtimedissipation}, with the appropriate rates adjusted according to the different confinement volumes considered. Note that for this confinement volume $S/V = 1/5d$ and that $\tau_V \approx 11T_D$, and that we use units normalized to the N-V center depth $d$ and diffusion time $T_D = d^2/D$, to display the universal behavior.} \label{DissipationPowers10}
\end{figure}

Spectroscopy demands stability of the correlation function, with the rate of decay of correlations typically controlling the minimum estimation error attainable \cite{Oviedo2020}. In Fig.\,\ref{DissipationPowers10}, we fix the parameters of the confinement volume to $R = L = 5d$, and study the influence of different dissipation rates, varying in powers of ten, from $\tev = T_D$ to $\tev = 10^6T_D$, to observe the onset of the $\textit{plateau}$ region in which the signal behaves coherently, and its disappearance due to dissipation. For slow dissipation rates $\tev \gg T_D$, the transition from diffusion to confinement and dissipation is well described by our model in Eq.\,\eqref{longtimedissipation} but, for dissipation rates which are on the order of the characteristic diffusion time, things are more complicated, as shown by the results for  $\tev = T_D$ and  $\tev = 10T_D$, for which the simple model for the long-time behavior of correlations represented by Eq.\,\eqref{longtimedissipation} overestimates the decay of correlations. The reason is that, when diffusion and dissipation have similar rates, their behavior cannot be as easily disentangled, as it is evident by the fact that the definition of $\tau_V$ is not valid anymore, and which also shows in the fact that no decay eigenvalue $\tau_{m,p}^\pm$ dominates significantly. Fast evaporation prevents the usefulness of confinement for the purpose of nano-NMR spectroscopy.

\begin{figure}
\includegraphics[width=\columnwidth]{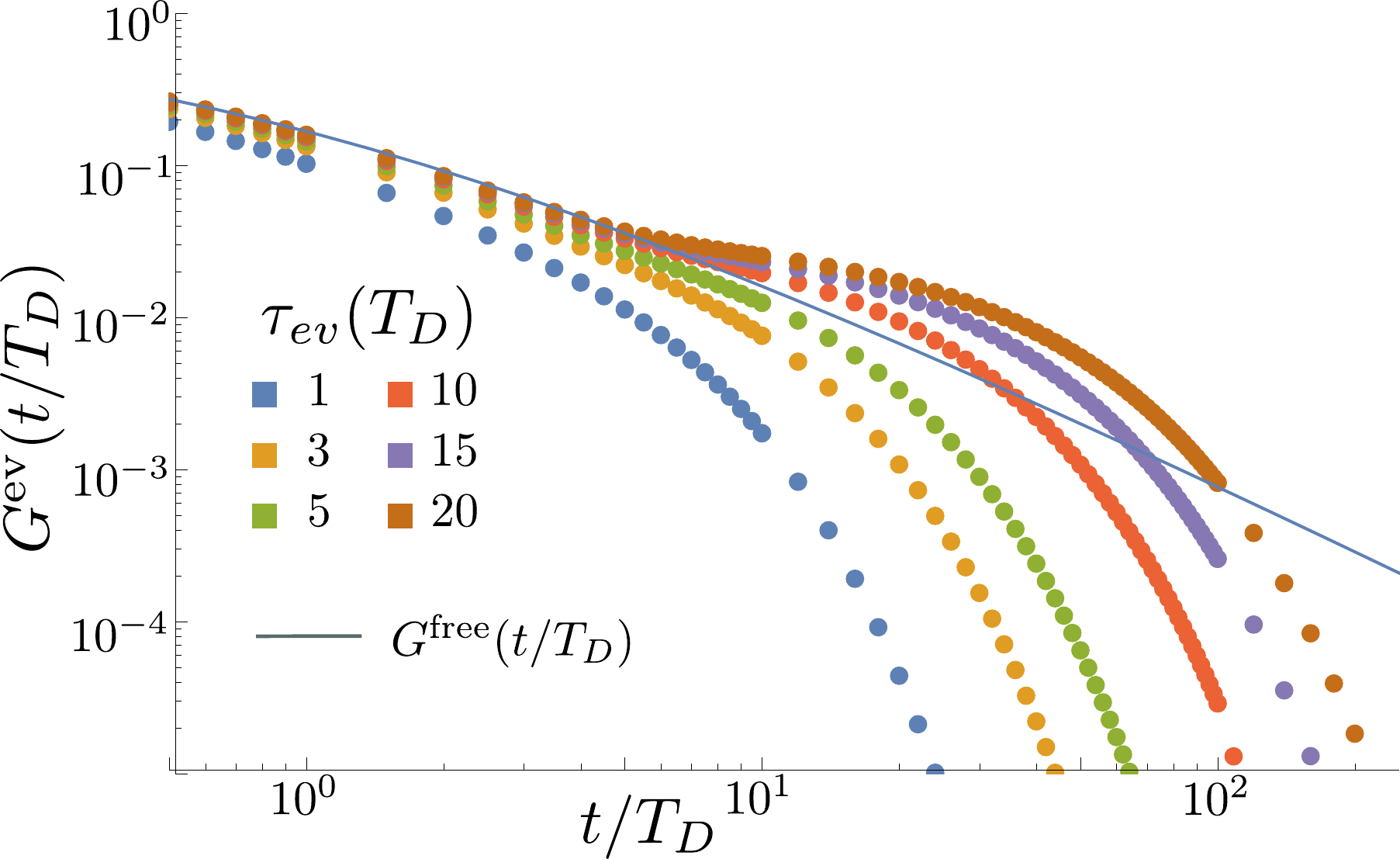}
    \caption{Logarithmic scale time evolution of the correlation function for confinement with fast evaporating walls, calculated from Eq.\,\eqref{evaporationcorrelation}. In this figure, the confinement cylinder volume features $R = 5d$ and $L = 5d$, and the evaporation rate ranges from $\tev = T_D$ to $\tev = 20T_D$. In gray, free diffusion correlation function  Eq.\,\eqref{Cohenian}. Here, we can no longer find a simple model that reproduces the behavior of the correlations with time, nor can we see a distinction between diffusive and evaporating regimes. Note that we use units normalized to the N-V center depth $d$ and diffusion time $T_D = d^2/D$, to display the universal behavior.} \label{DissipationFast}
\end{figure}

To explore the regime in which diffusion and dissipation are commensurate, we calculate, in Fig.\,\ref{DissipationFast}, the behavior of the correlation for fast dissipation rates, ranging from $\tev = T_D$ to $\tev = 20T_D$. We observe that, compared to the correlation function corresponding to free diffusion Eq.\,\eqref{Cohenian}, which accurately describes the initial time behavior when dissipation rates are slow, as shown in Fig.\,\ref{DissipationPowers10}, here, the decay of correlations that is induced by free diffusion underestimates the overall decay, even at early times. The intuitive reason is that, in this scenario, as soon as nuclei reach the walls of the confinement volume, they evaporate, and stop contributing to the overall magnetic field. The mathematical reason is that we no longer have a mode $\tau_{m,p}^\pm$ that dominates. Rather, many decay modes contribute on the same order of magnitude. Moreover, we no longer have a clear definition of $\tau_V$ that separates the diffusive and confinement regimes. Overall, the meaning is that in this scenario diffusion and evaporation mingle and act concurrently. In this case, a bigger confinement volume, such as those shown in Fig.\,\ref{DissipationVolumes}, in which the effect of the walls is delayed due to the nuclei taking longer to reach them, could help disentangling the mixture of evaporation and diffusion.

\section{Implications for frequency resolution}

The actual physics governing confined diffusion at the nanoscale, as displayed by the correlation function, is likely to be a combination of the two noise models described in this article, with the confining volume walls inducing a decay of the correlation function on some timescale $\tev$, as captured by the evaporating model, while at the same time slowing diffusion due to interaction with the nuclei in the sample, as described by the sticky model. The particular details of the exact model can be elucidated from experiments by comparing with the calculations shown above. In this section, we focus on a rather different implication of the realistic behavior of the correlation function of a confined sample. Namely, its usefulness for spectroscopy. In particular, we concentrate the ensuing discussion on frequency estimation. 

The ability to estimate a parameter from a signal depends on various things: the particular experiment, sensor, and protocol used to detect the signal, the underlying noise model, and the choice of data analysis. For nano-NMR spectroscopy with N-V centers, in which the statistical polarization is subjected to diffusion noise, the best known protocols are those of the phase-sensitive family \cite{Degen2017,Glenn2018}, such as Qdyne \cite{Schmitt2017}, which maximize frequency information extraction as benchmarked by the Fisher information \cite{Staudenmaier2023}.

According to the Crámer-Rao bound, the Fisher information for a given parameter determines the lower bound on the minimum error attainable when estimating that parameter from a given experiment. The Fisher information on a parameter $\delta$ given a discrete distribution $\{\tilde{P}_i(\delta)\}_i$ is calculated as 
\be
I_\delta = \sum_i\frac{(\frac{d\tilde{P}_i}{d\delta})^2}{\tilde{P}_i},
\ee
where in our case $\delta$ represents the (typically small) frequency to be estimated, and $\tilde{P}_i$ is the probability of the quantum sensor to be found at the state $i$, measured as the  average fluorescence. For a signal coming from statistically polarized diffusing nuclei, these probabilities directly relate to the auto-correlation of the magnetic field, such that
\be
\tilde{P} = \frac{1+ \frms^2\cos(\delta t)G(t/T_D)}{2},
\ee
is the probability of finding the probe in its original state, with $\frms \propto \brms$. For free diffusion, in which the correlation function envelope $G^{\text{free}}(t/T_D)$ in Eq.\,\eqref{Cohenian} follows a power-law at long times \cite{Cohen2019,Staudenmaier2022}, the Fisher information scaling with the total duration $T$ of an experiment is $I_\delta^{\text{free}} \propto T\log(\delta T)$ \cite{Oviedo2020}. Then, a sufficiently long total measurement time $T$ permits estimating small frequencies. On the other hand, an ideal confinement, with perfectly reflective walls or perfectly sticky walls, in which the correlation envelope reaches the \textit{plateau}, recovers the coherent signal limit, in which the oscillations do not decay \cite{Cohen2020}. Then, the Fisher information scales as $I_\delta \propto T^3$ \cite{Gefen2017}, albeit with a decreased amplitude which corresponds to the \textit{plateau} amplitude, which can be significantly smaller than the initial amplitude or $\brms$ [see Eq\,\eqref{eq:brms}] for large confinement regions. Consequently, either ideally reflective or sticky walls, combined with moderately sized confinement volumes where the \textit{plateau} amplitude is just one or two orders of magnitude smaller than the $\brms$ (see Fig.\,\ref{Stickycorrelation}), mean that using modest total measurement times $T$, a confinement spectroscopy experiment is a better strategy in terms of frequency estimation, particularly at low frequencies $\delta < \pi/T_D$. The Fisher information ratio between a sticky walls confinement and a free diffusion experiment is
\be
\frac{I_\delta^{\text{sticky}}}{I_\delta^{\text{free}}} \approx \frac{\left[G^{\text{sticky}}(t \gg \tau_V)\right]^4 T^2 }{\frms^4 T_D^2 \log{\delta T}},
\label{ratioinfo1}
\ee
where we assume the long experiment time limit $T \gg \tau_V$, and where for fully reflective walls  $G^{\text{ideal}}(t \gg \tau_V)$ replaces $G^{\text{sticky}}(t \gg \tau_V)$. Wisely choosing the confinement volume among those experimentally available, and finetuning the stickiness-to-reflectiveness ratio of the walls, allows to maximize information accumulation. As a general recipe, if walls can be made sticky, flatter geometries with $R \gg L$ are favored, whereas if only deep narrow wells are available for confinement, having reflective walls provides better conditions for spectroscopy.

Eq.\,\eqref{ratioinfo1} shows that free diffusion nano-NMR yields better frequency resolution only for very viscous fluids or exceedingly large confinement volumes. However, Eq.\,\eqref{ratioinfo1} disregards the possibility that the walls induce a decay of the correlations. Therefore, a more realistic comparison would be to consider a correlation function for the confinement regime which accounts for both sticky and evaporating walls, for which we can use the approximate expression Eq.\,\eqref{longtimedissipation}. In this case, the exact expression for the Fisher information is quite involved [see Appendix\,\ref{AppendixFI}], but can be studied in certain limits. In particular, we are interested in the limits of very fast and very slow evaporation.

When the walls induce a very fast decay of the correlations because $\tev\,\leq\,T_D$, the sample behaves, essentially, as a free diffusing fluid which suffers from an additional noise term rather than diffusion on the timescale of the evaporation time $\tev$, which causes added exponential correlation decay. Then, the Fisher information ratio between a confined and a free sample spectroscopy experiment is
\be
\frac{I_\delta^{\text{ev}}}{I_\delta^{\text{free}}} \approx \frac{\log\left(1+\tev^2\delta^2+\tev^4\delta^4\right)}{8 \log{\delta T}},
\label{ratioinfo2}
\ee
which explicitly demonstrates that, in this scenario, the best strategy is to not use confinement.

For large evaporation times $\tev \gg \left\{T_D,T_V\right\}$, it is instructive to explicitly show the approximate Fisher information expression for confined nano-NMR spectroscopy. To calculate it, we use the approximate model Eq.\,\eqref{longtimedissipation}, which yields:
\be
\begin{split}
I_\delta^{\text{ev}} &\approx \frac{\left[G^{\text{ideal}}(t \gg \tau_V)\right]^4\tilde\tau_{\text{ev}}^2}{16\tilde\tau^2} \\& 
\times\left[2\tilde\tau_{\text{ev}} T + 2T^2\exp\left(-\frac{2T}{\tilde\tau_{\text{ev}}}\right)  \right]\tanh{(\delta^2\tilde\tau_{\text{ev}}^2)},
\label{ficonfined}
\end{split}
\ee
with $\tilde\tau_{\text{ev}} = V\tev/Sd$. 

Eq.\,\eqref{ficonfined} features two distinct, important regimes, that depend on the specific target frequency $\delta$ and total measurement time $T$. In particular, Eq.\,\eqref{ficonfined} shows that the information about the frequency grows quadratically during the \textit{plateau} phase of the time evolution of the correlation function, becoming linear when $T > \tev$. As the hyperbolic tangent behaves as $\tanh{x} \sim x$ when $x \rightarrow 0$, this means that a resolution limit for small frequencies exists when $\delta < \pi/\tev$. On the other hand, $\tev$ can be substantially large, which not only provides a large measurement time $T$ of quadratic information accumulation but, notably, also means that even for quite small frequencies, sample confinement is a better strategy than having free diffusion, as the Fisher information comparison shows:
\be\label{eq:approxFIratio}
\frac{I_\delta^{\text{ev}}}{I_\delta^{\text{free}}} \approx \frac{\left[G^{\text{ideal}}(t \gg \tau_V)\right]^4\tev^2 T}{16\frms^4 T_D^3 \log{\delta T}}.
\ee
Since $\tev \gg T_D$ for good confinement, for frequencies $\delta \geq \pi/\tev$, confining the sample minimizes the frequency estimation error. For extremely small frequencies, Eq.\,\eqref{ficonfined} goes to zero, which means that the logarithmic behavior $\log(\delta T)$ of the Fisher information for free diffusion confers this strategy, in principle, with an advantage, despite requiring extremely long measurement times $T$ to reduce the estimation error. This can be inferred from Fig.\,\ref{DissipationPowers10}, where the gentler slope of the free diffusion correlation means that it eventually crosses with the faster decay of evaporation, albeit at a very long time and exceedingly small amplitude. Notice that, combining evaporating walls with some degree of stickiness in appropriate geometries can further enlarge the advantage for confinement, as the amplitude in Eq.\,\eqref{eq:approxFIratio} gets replaced by the sticky model amplitude in Eq.\,\eqref{eq:stickyplateau}, which can be significant particularly when the evaporation rate is large or target frequencies are small.

Summarizing, sample confinement provides an advantage for NMR spectroscopy of nanosized samples for fair sized confinement volumes on the order of $10^3 nm^3$, as long as the dephasing time due to interaction of the nuclei with the walls or due to seepage is large $\geq 1 s$. For very large confinement volumes $\sim 0.1 \mu m^3$ or fast dephasing times $\leq 100 \mu s$, an unconfined experiment has a larger chance of success. For medium confinement volumes or dephasing times, the experiment that yields the smallest estimation error can be predicted using the formulas calculated here.

\section{Conclusions}

Realizing sample confinement is a promising route to performing NMR spectroscopy on fluid samples with quantum sensors. A first experimental demonstration of actual confinement would require showing the characteristic \textit{plateau} that occurs when the diffusive regime transitions to the confined regime. In this article, we have modeled different confinement effects that can substantially modify the so-called ideal behavior of the correlations. By calculating the effect of having sticky walls, which slows diffusion close to the borders of the confinement region, we show how the amplitude of the \textit{plateau} can significantly change, not necessarily decreasing and, in fact, increasing in many geometries, and by calculating the effect of walls that induce dephasing or are porous, we demonstrate how the \textit{plateau} decays with time. A simplified model that combines both effects, when the evaporation time that controls the decay due to the walls is sufficiently different from the diffusion time, can be described by a simple expression that we provide, which can be used to predict the behavior of the correlations for a given experiment or to elucidate the properties of the experiment based on the observed correlations. Additionally, we provide exact expressions for estimating the diffusion coefficient, the expected \textit{plateau} amplitude and to calculate the frequency estimation error for precision spectroscopy.

Recently, a first attempt at experimental sample confinement was published by Liu et al.\,\cite{Liu2022}, where using a metal-organic framework they managed to slow diffusion to a trickle, greatly improving the frequency estimation ability by substantially reducing the width of the power spectrum of the sample. This seminal work demonstrates that metal-organic frameworks could be a nice strategy for sample confinement, as they can, essentially, be designed to have the desired mesh hole size, which means any confinement volume could be engineered. Our evaporation model describes accurately the effect of such a framework, where diffusion of molecules out of the confinement volume still happens, but at a much slower rate than that of free diffusion. Furthermore, this result has been followed, lately, by various successful attempts at partial confinement in assorted setups which demonstrate great improvement in slowing water diffusion, and uncover interesting behaviors regarding surface interactions or evaporation \cite{li2024,pagliero2025,zheng2026}, hinting at the significance of having reliable, accurate models capable of describing the unexpected behavior that confined liquids at the nanoscale can show.

The models presented here open up the pathway to understanding realistic sample confinement, and generate the foundation to study more complicated diffusion effects or to explore different confinement geometries through a mixture of analytical and numerical techniques. The interface between the diffusive and confinement regimes permits determining the diffusion coefficient of a fluid with improved accuracy, and an appropriate confinement can lead to enhanced resolution in nano-NMR spectroscopy.

\section{Acknowledgements}
We are grateful to Alex Retzker for fruitful discussions. S.O.-C acknowledges the financial support from the \textit{María Zambrano} Fellowship and from the Agencia Estatal de Investigación (QNAVIUM Project SCPP2400C011413XV0 funded by MICIU/AEI/10.13039/501100011033). D.C. acknowledges the
support of the Clore Scholars Programme and the Clore Israel Foundation. 
J.C. acknowledges support from grant PID2021-124965NB-C22 funded by MICIU/AEI/10.13039/501100011033 and by “ERDF/EU", European Union project C-QuENS (Grant No. 101135359), and grant CNS2023-144994 funded by MICIU/AEI/10.13039/501100011033 and by “ERDF/EU".

\section{Data availability}

The codes and data that support the findings in this article are openly available \cite{DataSharing}.

\appendix

\section{Diffusion on a cylinder with sticky walls}\label{AppendixSticky}

We aim to calculate the correlation function (in the time domain) for statistically polarized diffusing nuclei. Details on the derivation of the dephasing effect of the magnetic field that is produced by the nuclei at the N-V center position given a specific control protocol, and the specifics on how to derive expressions to calculate the corresponding correlation function, can be found in the Supplemental Material of Refs.\,\cite{Cohen2019,Cohen2020}. In this Appendix we focus on the derivation of the correlation function for the sticky walls model. 

The generic correlation function in the time domain reads
\be
G^{m_1,m_2}(t) \propto \int \int f^{m_1\,*}(\bar{r})f^{m_2}(\bar{r}_0)P(\bar{r},t|\bar{r}_0,t=0)d^3rd^3r_0, 
\label{correlationfunctionintegral}
\ee
where the $f^{m_i}(\barr\,)$ are the spatial dependencies of the respective magnetic fields generated by each pair of nuclei. Using the polar symmetry of the problem, given that we want to measure the correlations at the N-V center sensor position, and assuming that the N-V center is oriented in the $z$ direction, $f^{m_i}(\barr\,) = Y_2^{m_i}(\Omega)/\barr^{\,3}$ with $m = 0,1,2$. Each of these components can be probed separately, yielding similar results (namely and equal functional form up to a numerical factor). Therefore, throughout, we focus on the $m=0$ component and drop the superscript, for which 
\be
\label{eqSI:sphericalharmonic}
Y_2^0(\Omega) = \frac{4\sqrt{\pi}}{\sqrt{5}}(3\cdot\cos^2\theta - 1) = \frac{4\sqrt{\pi}}{\sqrt{5}}\left[\frac{3z^2}{(z^2 + \rho^2)}-1\right].
\ee

$P(\barr,t|\barro,t=0)$ in Eq.\,\eqref{correlationfunctionintegral} represents the nuclei propagator from point $\barro$ to $\barr$ in a time $t$. This propagator is the solution to a diffusion equation. Then, the first step is to solve that equation.

In this article, we consider confined Fickian diffusion without drift for the liquid sample, which corresponds to the motion of pressureless, compressible Navier-Stokes equations or, alternatively, to isothermal Navier-Stokes motion due to concentration gradients of an irrotational fluid \cite{Heifetz2022,Chanda2022}. The diffusion equation in three dimensions using cylindrical coordinates reads
\be
\frac{\partial P}{\partial t} -D\left[ \frac{1}{\rho}\frac{\partial }{\partial \rho}\left(\rho\frac{\partial P}{\partial \rho}\right) + \frac{1}{\rho^2}\frac{\partial ^2P}{\partial \phi^2} + \frac{\partial ^2P}{\partial z^2}\right]=0,
\ee
with $D$ the (constant) diffusion coefficient. The sticky model requires Dirichlet boundary conditions, which describe a vanishing propagator at the cylinder boundaries
\be
\begin{split}
&P(\barr,t|\barro,t=0) \at[\bigg]{\rho = R} = 0, \\&
 P(\barr,t|\barro,t=0)\at[\bigg]{z = d} = 0, \\&
P(\barr,t|\barro,t=0)\at[\bigg]{z = d+L} = 0. \\&
P(\phi+2\pi) = P(\phi).
\end{split}
\ee
Finally, the last ingredient to solve the diffusion equation are the initial conditions, which for us describe fixed starting points for the molecules
\be
P(\barr,t|\barro,t=0) = \frac{1}{\rho}\delta(\rho-\rho_0)\delta(z-z_0)\delta(\phi-\phi_0).
\ee

A general solution for this differential equation can be found in Refs.\,\cite{morse1953,carslaw1959,crank1975}, and the propagator can be obtained by adapting the generic solution found in Ref.\,\cite{Diffbook} to the particular initial and boundary conditions. Doing so yields 
\onecolumngrid
\be\label{eqSI:stickypropagator}
\begin{split}
    &P(\barr,t|\barro,t=0) = \frac{2}{V}\sum_{m=1}^\infty \sum_{n\in\mathrm{Z}}\sum_{p=0}^\infty \frac{J_n(\beta_{n,p}\rho)J_n(\beta_{n,p}\rho_0) \exp[-in(\phi-\phi_0)]}{J_{n+1} (\beta_{n,p}R)^2} \\& \times\left\{ \sin\left[\frac{2m\pi}{L}\left(z-d-\frac{L}{2}\right)\right]\sin\left[\frac{2m\pi}{L}\left(z_0-d-\frac{L}{2}\right)\right]  \exp\left[-D\left(\beta_{n,p}^2 + \frac{4m^2\pi^2}{L^2}\right)t\right]           \right. \\&  \left. + \cos\left[\frac{(2m-1)\pi}{L}\left(z-d-\frac{L}{2}\right)\right]\cos\left[\frac{(2m-1)\pi}{L}\left(z_0-d-\frac{L}{2}\right)\right] \exp\left[-D\left(\beta_{n,p}^2 + \frac{(2m-1)^2\pi^2}{L^2}\right)t\right]   \right\}.
\end{split}
\ee
The eigenvalues $\beta_{n,p}$ correspond to the solutions of the equation $J_n(\beta R) = 0$, where, for each $n$, there is a family $p = 0, 1, 2\dots$ of eigenvalues.

\onecolumngrid
\subsection{Bulk contribution to the correlation}

With the propagator in Eq.\,\eqref{eqSI:stickypropagator} derived above we can calculate the contribution to the correlation due to the nuclei that remain in the bulk of the cylinder at any given time through
\be
G^{\text{bulk}}(t) = \int_V \int_V \frac{Y_2^0(\Omega)}{\barr^{\,3}}\frac{Y_2^0(\Omega_0)}{\barro^{\,3}}P(\barr,t|\barro,t=0)drdr_0,
\ee
where using the symmetry in $\phi$, which yields $4\pi^4\delta_{n,0}$ we are left with 
\be
\begin{split}
G_{\text{bulk}}^{\text{sticky}}(t) &= \frac{8\pi^2}{V}\sum_{m=1}^\infty\sum_{p=0}^\infty \int_d^{d+L}dz\int_0^R d\rho \rho\int_d^{d+L}dz_0\int_0^R d\rho_0 \rho_0 \\& \left[\frac{3z^2 }{(\rho^2+z^2)^{5/2}} - \frac{1}{(\rho^2+z^2)^{3/2}} \right]\left[\frac{3z_0^2 }{(\rho_0^2+z_0^2)^{5/2}} - \frac{1}{(\rho_0^2+z_0^2)^{3/2}} \right]\frac{J_0(\beta_{0,p}\rho)J_n(\beta_{0,p}\rho_0) }{J_{1} (\beta_{0,p}R)^2} \\& \times\Bigg\{ \sin\left[\frac{2m\pi}{L}\left(z-d-\frac{L}{2}\right)\right]\sin\left[\frac{2m\pi}{L}\left(z_0-d-\frac{L}{2}\right)\right] \exp\left[-D\left(\beta_{0,p}^2 + \frac{4m^2\pi^2}{L^2}\right)t\right]          \\& +  \cos\left[\frac{(2m-1)\pi}{L}\left(z-d-\frac{L}{2}\right)\right]\cos\left[\frac{(2m-1)\pi}{L}\left(z_0-d-\frac{L}{2}\right)\right] \exp\left[-D\left(\beta_{0,p}^2 + \frac{(2m-1)^2\pi^2}{L^2}\right)t\right]   \Bigg\}.
\end{split}
\ee
Considering the symmetry that the integrals have allows to write the autocorrelation in the simplified form shown in Eq.\,\eqref{eq:bulkstickycorrelation} on the main text.

\subsection{Harmonic measure}

At large time, the contribution to the autocorrelation of the magnetic field comes from having all nuclei distributed on the surface of the cylinder. Such distribution is given by the harmonic measure, which corresponds to the outward derivative of the Green function in each of the surfaces of the cylinder. The Green function in turn is the time integral of the propagator. In mathematical terms: 
\be
\begin{split}
&G(t\gg\tau_v) = \int_V \frac{Y_2^0(\rho_0,z_0)}{\left(\rho_0^2+z_0^2\right)^{3/2}} \\& \left\{\int_S \left[\frac{\partial}{\partial n}D\int_0^\infty P(\barr,t|\barro,t=0)  dt \right]\frac{Y_2^0(\rho,z)}{\left(\rho^2+z^2\right)^{3/2}}dS\right\} dV.
\end{split}
\ee

If we consider the even and odd parts of the propagator, the Green function naturally splits as
\be
\tilde{G}(\barr,\barro) = \frac{P^{\text{even}}(\barr,t|\barro,t=0)}{\beta_{n,p}^2 + \frac{4m^2\pi^2}{L^2}} + \frac{P^{\text{odd}}(\barr,t|\barro,t=0)}{\beta_{n,p}^2 + \frac{(2m-1)^2\pi^2}{L^2}},
\ee
with the even (odd) components corresponding to the sine (cosine) of the $z$ eigenfunctions in the propagator. 

Now, the harmonic measure density is the (negative) outward derivative of the Green function in each of the surfaces. Beginning with the side wall at $\rho = R$ we have that the key derivative is
\be
-\frac{\partial J_n(\beta_{n,p}\rho)}{\partial \rho}\at[\bigg]{\rho = R} = -\frac{n}{R}J_n(\beta_{n,p}R) + \beta_{n,p}J_{n-1}(\beta_{n,p}R) = \beta_{n,p}J_{n+1}(\beta_{n,p}R),
\ee
where we use the Dirichlet boundary condition $J_n(\beta_{n,p}R) = 0$ and that when $J_n(\beta_{n,p}R) = 0$ then $J_{n-1}(\beta_{n,p}R) = -J_{n+1}(\beta_{n,p}R)$. The corresponding harmonic measure density at the radial wall of the cylinder is then 
\onecolumngrid
\be
\begin{split}
\omega^{\text{wall}}(R, z|\rho_0, z_0) &= \frac{2}{V}\sum_{m=1}^\infty \sum_{n\in\mathrm{Z}}\sum_{p=0}^\infty \frac{\beta_{n,p}J_n(\beta_{n,p}\rho_0) \exp[-in(\phi-\phi_0)]}{J_{n+1} (\beta_{n,p}R)}  \\& \times\left\{ \frac{1}{\beta_{n,p}^2 + \frac{4m^2\pi^2}{L^2}}\sin\left[\frac{2m\pi}{L}\left(z-d-\frac{L}{2}\right)\right] \sin\left[\frac{2m\pi}{L}\left(z_0-d-\frac{L}{2}\right)\right]           \right. \\& \left.  +  \frac{1}{\beta_{n,p}^2 + \frac{(2m-1)^2\pi^2}{L^2}}\cos\left[\frac{(2m-1)\pi}{L}\left(z-d-\frac{L}{2}\right)\right]  \cos\left[\frac{(2m-1)\pi}{L}\left(z_0-d-\frac{L}{2}\right)\right]   \right\}.
\end{split}
\ee

On the top and bottom caps the outward derivatives are, respectively, $\partial/\partial z$ at $z = d+L$ and $-\partial/\partial_z$ at $z = d$. Beginning with the top cap we have that:
\be
\frac{\partial}{\partial z} \sin\left[\frac{2m\pi}{L}\left(z-d-\frac{L}{2}\right)\right]\at[\bigg]{z = d+L} = \frac{2m\pi}{L}\cos(m\pi) = \frac{(-1)^m 2m\pi}{L},
\ee
and 
\be
\frac{\partial}{\partial z} \cos\left[\frac{(2m-1)\pi}{L}\left(z-d-\frac{L}{2}\right)\right]\at[\bigg]{z = d+L} = -\frac{(2m-1)\pi}{L}\sin\left[\frac{(2m-1)\pi}{2}\right] = \frac{(-1)^m(2m-1)\pi}{L},
\ee
while the bottom cap yields
\be
-\frac{\partial}{\partial z} \sin\left[\frac{2m\pi}{L}\left(z-d-\frac{L}{2}\right)\right]\at[\bigg]{z = d} = -\frac{2m\pi}{L}\cos(-m\pi) = -\frac{(-1)^m 2m\pi}{L},
\ee
and
\be
-\frac{\partial}{\partial z} \cos\left[\frac{(2m-1)\pi}{L}\left(z-d-\frac{L}{2}\right)\right]\at[\bigg]{z = d} = \frac{(2m-1)\pi}{L}\sin\left[-\frac{(2m-1)\pi}{2}\right] = -\frac{(-1)^m(2m-1)\pi}{L}.
\ee

The harmonic density of top and bottom caps is then
\be
\begin{split}
&\omega^{\text{bottom}}(\rho,d|\rho_0,z_0) =  -\omega^{\text{top}}(\rho,d+L|rho_0,z_0) = \frac{2}{V}\sum_{m=1}^\infty \sum_{n\in\mathrm{Z}}\sum_{p=0}^\infty \frac{J_n(\beta_{n,p}\rho)J_n(\beta_{n,p}\rho_0) \exp[-in(\phi-\phi_0)]}{J_{n+1} (\beta_{n,p}R)^2} \\& \times\left\{ \frac{(-1)^m 2m\pi}{L\left(\beta_{n,p}^2 + \frac{4m^2\pi^2}{L^2}\right)}\sin\left[\frac{2m\pi}{L}\left(z_0-d-\frac{L}{2}\right)\right] +             \frac{(-1)^m (2m-1)\pi}{L\left(\beta_{n,p}^2 + \frac{(2m-1)^2\pi^2}{L^2}\right)}\cos\left[\frac{(2m-1)\pi}{L}\left(z_0-d-\frac{L}{2}\right)\right]   \right\}.
\end{split}
\ee

Finally, the correlation at infinite time is the volume integral over the initial distribution of the harmonic density integral over the surface, with the corresponding dipole coupling terms. Then
\be
G(t\gg\tau_v) = \frac{1}{V}\int_V\int_S  \frac{Y_2^0(\rho,z)}{\left(\rho^2+z^2\right)^{3/2}}\frac{Y_2^0(\rho_0,z_0)}{\left(\rho_0^2+z_0^2\right)^{3/2}}\omega^S(\rho,z|\rho_0,z_0)dS dV.
\ee
Note that in all cases $\int_0^{2\pi} \exp[{-in(\phi-\phi_0)]}d\phi d\phi_0 = (2\pi)^2\delta_{n,0}$, which simplifies the sum in each harmonic density, and reduces the number of Bessel zeros to calculate. 

Changing the order of sums and integrals, and using the angular symmetry, we are left with
\be
\begin{split}
&G(t\gg\tau_V) = \frac{8\pi^2}{V}\sum_{m=1}^\infty \sum_{p=0}^\infty \int_d^{d+L}dz_0\int_0^R d\rho_0 \rho_0\left[\frac{3z_0^2 }{(\rho_0^2+z_0^2)^{5/2}} - \frac{1}{(\rho_0^2+z_0^2)^{3/2}} \right] \\& \times\left\{   \int_0^R d\rho \rho\, \left[\frac{3d^2 }{(\rho^2+d^2)^{5/2}} - \frac{1}{(\rho^2+d^2)^{3/2}} \right]\omega^{\text{bottom}}(\rho,d|\rho_0,z_0) \right.\\&\left.
+ \int_0^R d\rho \rho\, \left[\frac{3(d+L)^2 }{(\rho^2+(d+L)^2)^{5/2}} - \frac{1}{(\rho^2+(d+L)^2)^{3/2}} \right] \omega^{\text{top}}(\rho,d+L|\rho_0,z_0)\right.\\& \left.
+ R\int_d^{d+L} dz\, \left[\frac{3z^2 }{(R^2+z^2)^{5/2}} - \frac{1}{(R^2+z^2)^{3/2}} \right]  \omega^{wall}(R,z|\rho_0,z_0)  \right\},
\end{split}
\ee
where the harmonic density modes are, respectively:
\be
\begin{split}\label{eqSI:hdwalls}
\omega^{\text{wall}}(R,z|\rho_0,z_0) &= \frac{\beta_{0,p}J_0(\beta_{0,p}\rho_0)}{J_{1} (\beta_{0,p}R)}  \left\{ \frac{1}{\beta_{0,p}^2 + \frac{4m^2\pi^2}{L^2}}\sin\left[\frac{2m\pi}{L}\left(z-d-\frac{L}{2}\right)\right]\sin\left[\frac{2m\pi}{L}\left(z_0-d-\frac{L}{2}\right)\right]            \right. \\& \left.  +\frac{1}{\beta_{0,p}^2 + \frac{(2m-1)^2\pi^2}{L^2}}\cos\left[\frac{(2m-1)\pi}{L}\left(z-d-\frac{L}{2}\right)\right]\cos\left[\frac{(2m-1)\pi}{L}\left(z_0-d-\frac{L}{2}\right)\right]   \right\},
\end{split}
\ee
\be
\begin{split}\label{eqSI:hdcaps}
\omega^{\text{bottom}}(\rho,d|\rho_0,z_0) = -\omega^{\text{top}}(\rho,d+L|\rho_0,z_0) &=  \frac{J_0(\beta_{0,p}\rho)J_0(\beta_{0,p}\rho_0) }{J_{1} (\beta_{0,p}R)^2}  \Bigg\{ \frac{(-1)^m 2m\pi}{L\left(\beta_{0,p}^2 + \frac{4m^2\pi^2}{L^2}\right)}\sin\left[\frac{2m\pi}{L}\left(z_0-d-\frac{L}{2}\right)\right]             \\& +  \frac{(-1)^m (2m-1)\pi}{L\left(\beta_{0,p}^2 + \frac{(2m-1)^2\pi^2}{L^2}\right)}\cos\left[\frac{(2m-1)\pi}{L}\left(z_0-d-\frac{L}{2}\right)\right]   \Bigg\}.
\end{split}
\ee

\subsection{Time-dependent accumulation on the surface}

To complete the calculation we need the contribution from the surface as a function of time, for which we are required to calculate the distribution of nuclei on the surface as a function of time, which corresponds to the outward flux of nuclei at the cylinder surface. Then, at each time, we can integrate in time this outward flux, and integrate it over the surface with the dipolar couplings adn the angular component thus eliminated as in the previous calculations. Then
\be
G_{\text{surface}}^{\text{sticky}}(t) =  \int_V \frac{Y_2^0(\rho_0,z_0)}{\left(\rho_0^2 + z_0^2 \right)^{3/2}} \Bigg\{\int_S \frac{Y_2^0(\rho,z)}{\left(\rho^2 + z^2 \right)^{3/2}}  \times\left[-D\frac{\partial}{\partial n} \int_0^t P(\rho,z,t|\rho_0,z_0,t=0)  dt \right]dS\Bigg\} dV.
\ee

The time integral of the propagator can be easily calculated, as it behaves as follows:
If $P(\rho,z,t|\rho_0,z_0,t=0) = \sum_s P_s(\rho,z,t\rightarrow\infty|\rho_0,z_0,t=0)\exp(-\lambda_s t)$, yielding
\be 
\int_0^t P(\rho,z,t|\rho_0,z_0,t=0)  dt  = \sum_s \frac{P_s(\rho,z,t\rightarrow\infty|\rho_0,z_0,t=0)}{\lambda_s}\left(1-\exp(-\lambda_s t)\right).
\ee

Here, each of the components of the propagator at infinite time corresponds to the calculation above for the modes contributing to the harmonic measure, so that the time-dependent accumulation on the surface is just the mode by mode sum as the nuclei arrive to the surface. Considering that the propagator is separated into even and odd parts, we get that 
\be
\begin{split}
&G^{\text{surface}}(t) = -\frac{8\pi^2}{V}\sum_{m=1}^\infty \sum_{p=0}^\infty  \int_d^{d+L}dz_0\int_0^R d\rho_0 \rho_0\left[\frac{3z_0^2 }{(\rho_0^2+z_0^2)^{5/2}} - \frac{1}{(\rho_0^2+z_0^2)^{3/2}} \right] \\& \times\Bigg[ \left( 1-\exp\left[-D\left(\beta_{0,p}^2 + \frac{4m^2\pi^2}{L^2}\right)t\right]\right) \left\{   \int_0^R d\rho \rho\, \left[\frac{3d^2 }{(\rho^2+d^2)^{5/2}} - \frac{1}{(\rho^2+d^2)^{3/2}} \right] \omega_{\text{bottom}}^{\text{even}}(\rho,d|\rho_0,z_0) \right. \\& \left.
+ \int_0^R d\rho \rho\, \left[\frac{3(d+L)^2 }{(\rho^2+(d+L)^2)^{5/2}} - \frac{1}{(\rho^2+(d+L)^2)^{3/2}} \right]\omega_{\text{top}}^{\text{even}}(\rho,d+l|\rho_0,z_0) \right.\\& \left.
+ R\int_d^{d+L} dz\, \left[\frac{3z^2 }{(R^2+z^2)^{5/2}} - \frac{1}{(R^2+z^2)^{3/2}} \right]\omega_{wall}^{\text{even}}(R,z|\rho_0,z_0)    \right\} \\& + \left( 1-\exp\left[-D\left(\beta_{0,p}^2 + \frac{(2m-1)^2\pi^2}{L^2}\right)t\right]\right)  \left\{   \int_0^R d\rho \rho\, \left[\frac{3d^2 }{(\rho^2+d^2)^{5/2}} - \frac{1}{(\rho^2+d^2)^{3/2}} \right]\omega_{\text{bottom}}^{\text{odd}}(\rho,d|\rho_0,z_0)  \right. \\& \left.
+ \int_0^R d\rho \rho\, \left[\frac{3(d+L)^2 }{(\rho^2+(d+L)^2)^{5/2}} - \frac{1}{(\rho^2+(d+L)^2)^{3/2}} \right]\omega_{\text{top}}^{\text{odd}}(\rho,d+L|\rho_0,z_0) \right. \\& \left.
+ R\int_d^{d+L} dz\, \left[\frac{3z^2 }{(R^2+z^2)^{5/2}} - \frac{1}{(R^2+z^2)^{3/2}} \right]\omega_{wall}^{\text{odd}}(R,z|\rho_0,z_0)    \right\} \Bigg], 
\end{split}
\ee
where we define the even parts of the harmonic density those with the sine components in Eqs.\,\eqref{eqSI:hdwalls}\,and\,\eqref{eqSI:hdcaps}, and the odd those with cosine components.

\subsection{Initial correlation signal amplitude or \texorpdfstring{$\brms^2$}{brms2}}

For completeness, note that the $\brms$ corresponds to an initial uniform distribution on the cylinder volume, and it is therefore calculated as
\be
\begin{split}
&G(t=0) = \brms^2 \\&= \frac{16\pi}{5} \int_0^R d\rho \int_d^{d+L}dz \rho^2\left[\frac{3z^2 }{(\rho^2+z^2)^{5/2}} - \frac{1}{(\rho^2+z^2)^{3/2}} \right]^2.
\label{eq:brms}
\end{split}
\ee


\section{Diffusion in a cylinder with constant evaporation on the boundaries}\label{AppendixEvaporation}

In this section, we need the solution to the diffusion equation inside a cylinder region enclosed by walls that allow evaporation on their surface, reflected as Robin boundary conditions. As before, the solution can be adapted from \cite{Diffbook} for the appropriate boundary and initial conditions, or derived following Refs.\,\cite{morse1953,carslaw1959,crank1975}.

\be
\begin{split}\label{eqSI:evaporationpropagator}
P(\barr,t|\barro,t=0) = &\frac{L}{V}\sum_{m=1}^\infty\sum_{n=-\infty}^\infty\sum_{p=0}^\infty 
\frac{2 (\beta_{n,p} R)^2 J _n\left(\beta_{n,p}\rho_0\right)J_n(\beta_{n,p} \rho)}{L\left[ \left( \eta_0^2R^2+ (\beta_{n,p} R)^2-n^2 \right)J_n\left(\beta_{n,p} R \right)^2 \right]}\exp\left[ -in(\phi-\phi_0) - D\left(\beta_{n,p}^2+\eta_m^2\right) t \right]  \\ &  
\times\left\{\frac{\sin\left[\eta_m^- \left(z_0-d-\frac{L}{2}\right) \right]}{1-\sinc\left(\eta_m^- L\right)}\sin\left[\eta_m^- \left(z-d-\frac{L}{2}\right) \right] + \frac{\cos\left[\eta_m^+ \left(z_0-d-\frac{L}{2}\right) \right]}{1+\sinc\left(\eta_m^+ L\right)}\cos\left[\eta_m^+ \left(z-d-\frac{L}{2}\right) \right] \right\} .
\end{split}
\ee

Here, the axial eigenvalues $\eta_m^{\pm}$ are the solutions to the transcendental equation  \be
\left(\frac{\eta_0}{\eta^\pm}\right)^{\pm 1} = \pm \tan\frac{\eta^\pm L}{2},
\ee
with $\eta_0 = d/D\tev$,
and the radial eigenvalues $\beta_{n,p}$ correspond in this case to the positive solutions to the transcendental equation
\be
J_n'(\beta R) + \frac{d R}{\beta R D\tev}J_n(\beta R) = 0, 
\ee
with a family $p = 0, 1, 2, \dots$ for each $n$.

\subsection{Calculation of the auto-correlation}

Assuming only one of the dipole-dipole interaction terms is on-resonance the auto-correlation reads,
\be
G^{(k)}(t)=\int \frac{d^3r}{r^3}\int \frac{d^3r_0}{r_0^3} Y_2^{(k)}(\vec{r}_0)\left[Y_2^{(k)}(\vec{r})\right]^* P(\vec{r},\vec{r}_0,t).
\ee
The angular integration over $\phi$ and $\phi_0$ yields $(2\pi)^2\delta_{k,n}$, therefore, using the symmetry of the integrals over $\barr,\barro$ the autocorrelation can be written as
\be
\begin{split}
G_k^{\text{ev}}(t) &= \frac{8\pi^2}{V}\sum_{m=1}^\infty\sum_{p=0}^\infty \frac{(\beta_{k,p}R)^2}{\left[(\eta_0R)^2+(\beta_{k,p}R)^2-k^2\right]J_k^2\left(\beta_{k,p}R\right)} \\& \times 
\Bigg( \frac{\exp\left[-t/D(\beta_{k,p}^2+{\eta_m^-}^2)\right]}{1-\sinc\left(\eta_m^- L\right)}\Bigg\{ \int_d^{d+L} dz \int_0^R d\rho \frac{\rho Y_2^k(z,\rho)J_k\left(\beta_{k,p}\rho\right)}{\left(\rho^2 + z^2 \right)^{3/2}}   \sin\left[\eta_m^- \left(z-d-\frac{L}{2}\right) \right] \Bigg\}^2  \\&  
+ \frac{\exp\left[-t/D(\beta_{k,p}^2+{\eta_m^+}^2)\right]}{1+\sinc\left(\eta_m^+ L\right)}\Bigg\{ \int_d^{d+L} dz \int_0^R d\rho \frac{\rho Y_2^k(z,\rho)J_k\left(\beta_{k,p}\rho\right)}{\left(\rho^2 + z^2 \right)^{3/2}}  \cos\left[\eta_m^+ \left(z-d-\frac{L}{2}\right) \right] \Bigg\}^2\Bigg),
\end{split}
\ee
where we can define the corresponding decay coefficients as $\tau_{k,m,p}^{\pm} = 1/D(\beta_{k,p}^2+{\eta_m^\pm}^2)$. In this article, $k = 0$, yielding Eq.\,\eqref{evaporationcorrelation}. 
\twocolumngrid

\section{Details for Fisher information calculations}\label{AppendixFI}

The Fisher information is an objective measure of the maximum accuracy that a statistical estimator can achieve about a given parameter in a specified experimental setting. According to the Cramér-Rao bound, the Fisher information provides a lower bound on the minimum mean-squared-error that can be achieved for an estimator of the target parameter. For this reason, it can be used to compare the expected performance of two different experimental protocols with the same target parameter. Here, we detail the calculations that lead to the approximate Fisher information ratios shown in the main text. Particular details on the Fisher information for frequency estimation with a coherent signal can be found in \cite{Schmitt2017,LowRamsey}, while for the free diffusion case the calculations are detailed in \cite{Oviedo2020,Staudenmaier2023}.

Our starting point is the probability to find the qubit on its upper (lower) state following some evolution time of an initial equal superposition state subjected to the noisy magnetic signal created by a distribution of diffusing nuclei. We assume a single target frequency $\delta$, for which the magnetic signal is described as $B(t) = a(t)\cos(\delta t) + b(t)\sin(\delta t)$, where the amplitudes $\{a(t),b(t)\}$ follow some distribution due to noisy fluctuations. As a result of the interaction with the signal, the qubit probe acquires a phase that depends on $B(t)$ and the particular dynamical decoupling sequence employed to control the qubit, which is modeled through the response function $h(t)$. This phase can be calculated as
\be
\Phi[B(t')] = \gamma_n \int_{-\tau/2}^{\tau/2} ds\, h(s)B(t'+s),
\ee
with $\tau$ the sequence duration and $\gamma_n$ is the gyromagnetic ratio of the nuclei. If we assume that $\tau$ is much smaller than any noise characteristic time affecting the system or the sample, and we consider that the pulses have negligible duration, the response function simplifies to $h(t') = \theta(t'+\tau/2)\theta(t'-\tau/2)$. 

In Qdyne, after every phase acquisition period, an interrogation of the qubit state is performed. The probability to find the qubit in its upper state at that moment is $P \propto \sin(\Phi_t)/2$. The sequence of phase acquisition plus interrogation is repeated in a synchronized fashion throughout the experiment, keeping $\tau$ constant, and controlling the interrogation and reinitialization time of the N-V center --the overhead time $\tau_o$-- constant as well. Accurately tracking the time $\tilde\tau = \tau + \tau_o$ enables reconstruction of the signal during the postprocessing, and allows us to correlate every pair of measurements, regardless of their temporal separation. Then, the relevant probability used to calculate the Fisher information is that of obtaining a pair of correlated measurements separated by an arbitrary time, which corresponds to the covariance between measurements. Assuming the weak signal limit, this covariance reads $\langle P_0 P_{j\tilde\tau}\rangle \sim \frms^2 \cos(\delta j\tilde\tau) G(j\tilde\tau )$, with $j \in \mathrm{N}$ representing any given measurement. The Fisher information for one such correlated pair then reads $i_{\delta} \propto \frms^4 t^2 \sin ^2 (\delta t ) G^2( t )$, and the total Fisher information $I_\delta$ is the sum over all possible correlated pairs of measurements within the total experiment time. Considering that a single measurement has a duration $\tilde\tau$, we get that 
\be
I_\delta \propto \frms^4\sum_{j=0}^{T/\tilde\tau}  \left(\frac{T}{\tilde\tau}-j\right)(j\tilde\tau)^2 \sin ^2 (\delta j \tilde\tau ) [G( j\tilde\tau)]^2,
\label{eq:generic_sum}
\ee
where the extra factor $(T/\tilde\tau - j)$ accounts for all possible subsequent measurements to which a given measurement can be correlated.

In the case of sticky walls, the distinct behavior from free diffusion is achieved when the autocorrelation reaches the \textit{plateau} section and stops decaying. At that point, the oscillating magnetic field behaves as a coherent signal with amplitude $\left[G^{\text{sticky}}(t \gg \tau_V)\right]^2$, meaning that the sum in Eq.\,\eqref{eq:generic_sum} can be calculated as
\be
\begin{split}
&I_\delta \propto \left[G^{\text{sticky}}(t \gg \tau_V)\right]^4 \sum_{j=0}^{T/\tilde\tau}  \left(\frac{T}{\tilde\tau}-j\right)(j\tilde\tau)^2 \sin ^2 (\delta j \tilde\tau ) \\& \approx \frac{\left[G^{\text{sticky}}(t \gg \tau_V)\right]^4 T_D T^3}{\tilde\tau^2},
\end{split}
\ee
where we just keep the leading order in $T$, and assume that $\delta T \gg 1$. Thus we get Eq.\,\eqref{ratioinfo1} in the main text.

The case of evaporating walls is slightly more complicated, as the sum in Eq.\,\eqref{eq:generic_sum} cannot be calculated directly. Rather, we have to calculate it as an integral using the Riemann recipe. Additionally, rather than trying to calculate directly with the correlation function envelope Eq.\,\eqref{evaporationcorrelation}, we use the various approximations depending on the different decay timescales. Thus, for very fast evaporation $\tev \leq T_D$ we have that the correlations decay, approximately, as $\sim t^{-3/2}\exp{(-t/\tev)}$, and the Fisher information is
\be
\begin{split}
I_\delta^{\text{ev}} &\sim \frac{\frms^4 T_D^4}{\tilde\tau^2}\int_0^{T/T_D} dz \left(\frac{T}{T_D}-z\right)z^2\sin^2(\delta z T_D) \\& \times\left[z^{-3}\exp{\left(-\frac{2 z T_D}{\tev}\right)}\right], 
\end{split}
\ee
where we use $z = t/T_D$ for simplicity. Assuming that $\delta T \gg 1$  and that $\delta\tev < 1$ this integral yields
\be
I_\delta^{\text{ev}} \approx \frac{\frms^4 T_D^4 T}{8\tilde\tau^2}\log\left( 1 + \delta^2\tev^2+ \delta^4\tev^4 \right),
\ee
with which we obtain Eq.\,\eqref{ratioinfo2} in the main text.

On the opposite end, when the evaporation timescale is much longer than diffusion or the volume time $\tev \gg \left\{T_D,T_V\right\}$, we can use the simplified model 
\be
\begin{split}
&G^{\text{ev}}(t) \approx \brms^2G^{\text{free}}(t/T_D)\exp(-2t/\tau_V) \\& + G^{\text{ideal}}(t \gg \tau_V )\exp\left(-\frac{t}{\tilde\tau_{\text{ev}}}\right),
\end{split}
\ee
to calculate the Fisher information, where $\tilde\tau_{\text{ev}} = V\tev/Sd$. Performing the same sum to integral transformation as before we get that 
\onecolumngrid
\be
\begin{split}
I_\delta^{\text{ev}} &\approx \frac{\brms^4T_D^4\tau_V^3\delta^2}{4\tilde\tau^2(T_D+T_D\tau_V^3\delta^2)} 
+\frac{G^{\text{ideal}}(t \gg \tau_V )^4 \tilde\tau_{\text{ev}} T \left[\tilde\tau_{\text{ev}}^2-2T^2\exp\left(\frac{-2T}{\tilde\tau_{\text{ev}}}\right)\right]}{8\tilde\tau^2}  \\&
+\frac{G^{\text{ideal}}(t \gg \tau_V )^4 \tilde\tau_{\text{ev}}}{16\tilde\tau^2}\left[ 3\tilde\tau_{\text{ev}}^3  - \exp\left(\frac{-2T}{\tilde\tau_{\text{ev}}} \right)\left( 4T^3+6T^2\tilde\tau_{\text{ev}}+6T\tilde\tau_{\text{ev}}^2+3\tilde\tau_{\text{ev}}^3\right)\right] \\& 
+\frac{\brms^2G^{\text{ideal}}(t \gg \tau_V )^2\tilde\tau_{\text{ev}}\tau_V}{4\sqrt{T_D}(\tilde\tau_{\text{ev}}+\tau_V)^2}  \times\left\{2\sqrt{T}\left[3\tilde\tau_{\text{ev}}\tau_V+2T\left(\tilde\tau_{\text{ev}} + \tau_V \right)\right]\exp\left(-\frac{T}{\tilde\tau_{\text{ev}}}--\frac{T}{\tau_V}\right)  -3\sqrt{\pi}\tilde\tau_{\text{ev}}\tau_V\sqrt{\frac{\tilde\tau_{\text{ev}}}{\tilde\tau_{\text{ev}}+\tau_V}}\right\}\\&
+\frac{\brms^2G^{\text{ideal}}(t \gg \tau_V )^2 T\tilde\tau_{\text{ev}}\tau_V\sqrt\pi }{2\sqrt{T_D}(\tilde\tau_{\text{ev}}+\tau_V)}\sqrt{\frac{\tilde\tau_{\text{ev}}}{\tilde\tau_{\text{ev}}+\tau_V}}
+\frac{\brms^2T_DT}{8}\log\left(1+2\tilde\tau_{\text{ev}}\delta^2 + \tilde\tau_{\text{ev}}^4\delta^4 \right).
\label{eq:evcompletefi}
\end{split}
\ee

Considering the limit of large experiment time $T\gg\tev$, as well as $\delta T \gg 1$  and $\delta\tev < 1$ we get that the dominant term reads
\be
I_\delta^{\text{ev}} \approx \frac{\left[G^{\text{ideal}}(t \gg \tau_V)\right]^4\tilde\tau_{\text{ev}}^2}{16\tilde\tau^2}\left[2\tilde\tau_{\text{ev}} T + 2T^2\exp(-\frac{2T}{\tilde\tau_{\text{ev}}})  \right]\tanh{(\delta^2\tilde\tau_{\text{ev}}^2)},
\ee
which corresponds to Eq.\,\eqref{ficonfined} in the main text.

\twocolumngrid



%

\end{document}